\documentclass[11pt,a4paper]{article}

\usepackage[margin=0.9in]{geometry}
\usepackage[T1]{fontenc}
\usepackage[utf8]{inputenc}
\usepackage{mathptmx}
\usepackage{lmodern}
\usepackage{microtype}
\usepackage{setspace}
\usepackage[inline,shortlabels]{enumitem}
\usepackage[super,sort,compress]{cite}
\usepackage{amsmath,amssymb,amsthm}
\usepackage{graphicx}
\usepackage{booktabs,array,threeparttable,multicol}
\usepackage[font=small,labelfont=bf]{caption}
\usepackage{xspace,xcolor}
\usepackage{ragged2e}
\usepackage{url}
\usepackage[colorlinks=true,linkcolor=blue,citecolor=blue,urlcolor=red]{hyperref}
\usepackage{nameref}
\usepackage[numbers,sort&compress]{natbib}
\usepackage{newunicodechar}
\newunicodechar{−}{-}
\newunicodechar{́}{\'{} }

\newcommand{\kms}{\mathrm{km\,s^{-1}}}
\newcommand{\msun}{M_{\odot}}

\newcommand{\hi}{H{\,\textsc{i}}\xspace}
\newcommand{\hj}{\text{\hi}}

\newcommand{\araa}{Annu. Rev. Astron. Astrophys.}   
\newcommand{\aj}{Astron. J.}   
\newcommand{\apj}{Astrophys. J.}   
\newcommand{\apjl}{Astrophys. J. Lett.}   
\newcommand{\apjs}{Astrophys. J. Suppl. Ser.}   
\newcommand{\aap}{Astron. Astrophys.}   
\newcommand{\mnras}{Mon. Not. R. Astron. Soc.}   
\newcommand{\nat}{Nature} 
\newcommand{\pasp}{Publ. Astron. Soc. Pac.}   
\newcommand{\dotmin}{\rlap.{'}}
\newcommand{\dotdeg}{\rlap.{^\circ}}

\newcommand{\gh}[1]{{\leavevmode\color{black} #1}}

\makeatletter
\renewcommand{\maketitle}{%
  \begin{center}
  {\LARGE\bfseries \@title\par}
  \vspace{0.7em}
  {\small \@author\par}
  \vspace{0.7em}
  {\footnotesize \@date\par}
  \end{center}
  \vspace{1em}
}
\makeatother

\usepackage[symbol]{footmisc}

\title{\textbf{Weak Evolution of Cosmic Atomic Hydrogen over the Past 4.5 Billion Years}}

\author{
Chuan-Peng~Zhang$^{1,5}$\textsuperscript{\textdagger},
Hong~Guo$^{2}$\textsuperscript{\textdagger,\textdaggerdbl},
Yizhou~Gu$^{3,4}$\textsuperscript{\textdagger},
Amélie~Saintonge$^{6,7}$,
Xiaohu~Yang$^{3,4}$,
Dirk~Scholte$^{8}$,
Ming~Zhu$^{1,5}$\textsuperscript{\textdaggerdbl},
Peng~Jiang$^{1,5}$\textsuperscript{\textdaggerdbl},
Hu~Zou$^{1}$,
Manasvee~Saraf$^{6}$,
Wenlin~Ma$^{2,9}$,
Yirong~Wang$^{4}$,
Y.~P.~Jing$^{4,3}$,
Zheng~Zheng$^{10}$,
Zhejie~Ding$^{11}$, 
J.~Aguilar$^{12}$,
S.~Ahlen$^{13}$,
D.~Bianchi$^{14,15}$,
D.~Brooks$^{6}$,
T.~Claybaugh$^{12}$,
A.~de la Macorra$^{16}$,
P.~Doel$^{6}$,
E.~Gaztañaga$^{17,18,19}$,
G.~Gutierrez$^{20}$,
M.~Ishak$^{21}$,
R.~Joyce$^{22}$,
S.~Juneau$^{22}$,
R.~Kehoe$^{23}$,
D.~Kirkby$^{24}$,
A.~Kremin$^{12}$,
O.~Lahav$^{6}$,
C.~Lamman$^{25}$,
M.~Landriau$^{12}$,
L.~Le~Guillou$^{26}$,
M.~Manera$^{27,28}$,
A.~Meisner$^{22}$,
R.~Miquel$^{29,28}$,
J.~Moustakas$^{30}$,
S.~Nadathur$^{18}$,
W.~J.~Percival$^{31,32,33}$,
F.~Prada$^{34}$,
I.~P\'erez-R\`afols$^{35}$,
G.~Rossi$^{36}$,
E.~Sanchez$^{37}$,
D.~Schlegel$^{12}$,
M.~Schubnell$^{38}$,
H.~Seo$^{39}$,
J.~Silber$^{12}$,
D.~Sprayberry$^{22}$,
G.~Tarl\'{e}$^{38}$,
B.~A.~Weaver$^{22}$
and FASHI Collaboration.\\
}

\date{\footnotesize  
$^{1}$ State Key Laboratory of Radio Astronomy and Technology, National Astronomical Observatories, Chinese Academy of Sciences, Beijing 100101, China\\
$^{2}$ Shanghai Astronomical Observatory, Chinese Academy of Sciences, Nandan Road 80, Shanghai 200030, China\\
$^{3}$ Tsung-Dao Lee Institute and Key Laboratory for Particle Physics, Astrophysics and Cosmology, Ministry of Education, Shanghai Jiao Tong University, Shanghai 201210, China\\
$^{4}$ Department of Astronomy, School of Physics and Astronomy, and Shanghai Key Laboratory for Particle Physics and Cosmology, Shanghai Jiao Tong University, Shanghai 200240, China\\
$^{5}$ Guizhou Radio Astronomical Observatory, Guizhou University, Guiyang 550000, China\\
$^{6}$ Department of Physics \& Astronomy, University College London, Gower Street, London, WC1E 6BT, UK\\
$^{7}$ Max Planck Institute for Radio Astronomy, Auf dem H\"ugel 69, 53121 Bonn, Germany\\
$^{8}$ Institute for Astronomy, University of Edinburgh, Royal Observatory, Blackford Hill, Edinburgh EH9 3HJ, UK\\
$^{9}$ University of Chinese Academy of Sciences, Beijing 100049, China\\
$^{10}$ Department of Physics and Astronomy, The University of Utah, 115 South 1400 East, Salt Lake City, UT 84112, USA\\
$^{11}$ University of Chinese Academy of Sciences, Nanjing 211135, China\\
$^{12}$ Lawrence Berkeley National Laboratory, 1 Cyclotron Road, Berkeley, CA 94720, USA\\
$^{13}$ Department of Physics, Boston University, 590 Commonwealth Avenue, Boston, MA 02215 USA\\
$^{14}$ Dipartimento di Fisica ``Aldo Pontremoli'', Universit\`a degli Studi di Milano, Via Celoria 16, I-20133 Milano, Italy\\
$^{15}$ INAF-Osservatorio Astronomico di Brera, Via Brera 28, 20122 Milano, Italy\\
$^{16}$ Instituto de F\'{\i}sica, Universidad Nacional Aut\'{o}noma de M\'{e}xico,  Circuito de la Investigaci\'{o}n Cient\'{\i}fica, Ciudad Universitaria, Cd. de M\'{e}xico  C.~P.~04510,  M\'{e}xico\\
$^{17}$ Institut d'Estudis Espacials de Catalunya (IEEC), c/ Esteve Terradas 1, Edifici RDIT, Campus PMT-UPC, 08860 Castelldefels, Spain\\
$^{18}$ Institute of Cosmology and Gravitation, University of Portsmouth, Dennis Sciama Building, Portsmouth, PO1 3FX, UK\\
$^{19}$ Institute of Space Sciences, ICE-CSIC, Campus UAB, Carrer de Can Magrans s/n, 08913 Bellaterra, Barcelona, Spain\\
$^{20}$ Fermi National Accelerator Laboratory, PO Box 500, Batavia, IL 60510, USA\\
$^{21}$ Department of Physics, The University of Texas at Dallas, 800 W. Campbell Rd., Richardson, TX 75080, USA\\
$^{22}$ NSF NOIRLab, 950 N. Cherry Ave., Tucson, AZ 85719, USA\\
$^{23}$ Department of Physics, Southern Methodist University, 3215 Daniel Avenue, Dallas, TX 75275, USA\\
$^{24}$ Department of Physics and Astronomy, University of California, Irvine, 92697, USA\\
$^{25}$ The Ohio State University, Columbus, 43210 OH, USA\\
$^{26}$ Sorbonne Universit\'{e}, CNRS/IN2P3, Laboratoire de Physique Nucl\'{e}aire et de Hautes Energies (LPNHE), FR-75005 Paris, France\\
$^{27}$ Departament de F\'{i}sica, Serra H\'{u}nter, Universitat Aut\`{o}noma de Barcelona, 08193 Bellaterra (Barcelona), Spain\\
$^{28}$ Institut de F\'{i}sica d’Altes Energies (IFAE), The Barcelona Institute of Science and Technology, Edifici Cn, Campus UAB, 08193, Bellaterra (Barcelona), Spain\\
$^{29}$ Instituci\'{o} Catalana de Recerca i Estudis Avan\c{c}ats, Passeig de Llu\'{\i}s Companys, 23, 08010 Barcelona, Spain\\
$^{30}$ Department of Physics and Astronomy, Siena University, 515 Loudon Road, Loudonville, NY 12211, USA\\
$^{31}$ Department of Physics and Astronomy, University of Waterloo, 200 University Ave W, Waterloo, ON N2L 3G1, Canada\\
$^{32}$ Perimeter Institute for Theoretical Physics, 31 Caroline St. North, Waterloo, ON N2L 2Y5, Canada\\
$^{33}$ Waterloo Centre for Astrophysics, University of Waterloo, 200 University Ave W, Waterloo, ON N2L 3G1, Canada\\
$^{34}$ Instituto de Astrof\'{i}sica de Andaluc\'{i}a (CSIC), Glorieta de la Astronom\'{i}a, s/n, E-18008 Granada, Spain\\
$^{35}$ Departament de F\'isica, EEBE, Universitat Polit\`ecnica de Catalunya, c/Eduard Maristany 10, 08930 Barcelona, Spain\\
$^{36}$ Department of Physics and Astronomy, Sejong University, 209 Neungdong-ro, Gwangjin-gu, Seoul 05006, Republic of Korea\\
$^{37}$ CIEMAT, Avenida Complutense 40, E-28040 Madrid, Spain\\
$^{38}$ Department of Physics, University of Michigan, Ann Arbor, MI 48109, USA\\
$^{39}$ Department of Physics \& Astronomy, Ohio University, 139 University Terrace, Athens, OH 45701, USA\\
\textsuperscript{\textdagger}{These authors contributed equally to this work and are co-first authors.}\\
\textsuperscript{\textdaggerdbl}{ Corresponding Authors.}
}

\begin{document}
\maketitle

\begin{abstract}
The cosmic star formation rate density (CSFRD) has declined sharply toward the present day, but the roles of the atomic and molecular gas reservoirs remain uncertain. We measure the cosmic \hi density, $\Omega_\hj$, over $0<z<0.41$ by combining \hi spectra from the Five-hundred-meter Aperture Spherical Telescope with optical spectroscopy from the Dark Energy Spectroscopic Instrument for $\sim2.5$ million galaxies across $\sim12,000\,{\rm deg}^2$. We measure a raw decrease in $\Omega_{\mathrm{HI}}$ by a factor of $1.35\pm0.10$ over the past 4.5 Gyr. Even after applying the conservative systematic corrections from our forward model, the inferred decline is only $1.12\pm0.10$ --- still far weaker than the CSFRD decline (a factor of 2.46). The molecular gas density, in contrast, is known to evolve more closely with star formation. At fixed stellar mass, the average \hi gas fraction evolves by less than 0.2 dex, showing that the weak evolution is present across the galaxy population. These quantitative differences rule out rapid depletion of galaxy \hi as the primary driver of the late-time CSFRD decline, and provide a stringent benchmark for models of gas accretion, phase conversion and star-formation regulation.
\end{abstract}

\section{Main}
Star formation in galaxies proceeds through the baryon cycle, in which gas accreted from the intergalactic medium (IGM) and circumgalactic medium (CGM) is transformed into atomic hydrogen (\hi), molecular hydrogen (H$_2$) and eventually stars \cite{Madau2014,Peroux2020}. The late-time decline of the cosmic star formation rate density (CSFRD) is widely linked to the observed decrease in the cosmic H$_2$ density, yet the role of the more extended \hi reservoir remains less clear \cite{Tacconi2020,Saintonge2022}. Previous measurements have provided important evidence that the cosmic \hi abundance, $\Omega_\hj$, evolves more mildly than the CSFRD. In particular, damped Ly$\alpha$ (DLA) absorbers provide essential constraints on the cosmic \hi budget over a broad redshift range, while blind and stacked 21-cm surveys measure \hi emission associated with galaxies at low and intermediate redshifts\cite{Peroux2020,Walter2020}. These approaches are complementary and together define the current observational picture. However, they are not directly interchangeable: DLA measurements are primarily sensitive to high-column-density neutral gas cross sections, whereas 21-cm emission measurements trace the integrated \hi mass associated with galaxies. Thus, they probe the cosmic \hi population through different selection functions and systematic uncertainties. Self-consistent late-time 21-cm emission measurements are still essential for testing whether the CSFRD decline is accompanied by comparable depletion of the galaxy-associated \hi reservoir.

While $\Omega_\hj$ in the local Universe is well characterized by blind 21-cm surveys \cite{Zwaan2005,Martin2010,Jones2018,Xi2021,Guo2023,Ponomareva2023,Ma2025}, extending such measurements to higher redshifts is challenging because of the intrinsic faintness of the 21-cm line. Existing intermediate-redshift 21-cm studies have therefore relied primarily on spectral line stacking \cite{Lah2007,Delhaize2013,Rhee2013,Rhee2016,Rhee2018,Bera2019,Hu2019,Chen2021,Rhee2023}, but face a fundamental observational trade-off: wide-area surveys lack the sensitivity to detect faint \hi emission at cosmological distances, whereas deep pencil-beam observations probe volumes too small to overcome cosmic variance. As a result, existing emission measurements at $z > 0.1$ are based on samples spanning only a few square degrees and at most a few hundred galaxies \cite{Lah2007,Rhee2018,Bera2019,Chowdhury2020}, and are subject to low signal-to-noise ratios, substantial cosmic variance, heterogeneous optical selections (see Extended Data Table~\ref{tab:HI_measurements} in \hyperref[sec:Methods]{Methods}). Resolving the role of \hi in the late-time decline of CSFRD therefore requires observations that are simultaneously deep, wide, statistically representative, and accompanied by well-characterized systematics.

\subsubsection{HI stacking observations}
We exploit a unique optical-radio synergy that combines the sensitivity of the Five-hundred-meter Aperture Spherical radio Telescope (FAST) with the survey scale and redshift precision of the Dark Energy Spectroscopic Instrument (DESI). Among existing blind 21-cm surveys, the FAST All-Sky \hi survey (FASHI) provides the deepest single-dish observations over a wide frequency range\cite{Zhang2024}, enabling sensitive measurements of atomic gas in galaxies out to intermediate redshifts. Cross-matching FASHI with DESI spectroscopy yields accurate redshifts over an unprecedented survey volume, producing a statistically representative sample of $\sim2.5$ million galaxies covering $\sim12,000\deg^2$ (see Extended Data Figure\,\ref{fig:observed_sky}). This combination makes it possible to measure $\Omega_\hj$ with substantially improved statistical precision over $0<z<0.41$, corresponding to the past 4.5 billion years.

We measure the atomic gas content using \hi spectral stacking (see stacked spectra in Extended Data Figures~\ref{fig:spec_z0.05}, \ref{fig:spec_z0.1}, and \ref{fig:spec_z0.3}) of r-band magnitude-limited galaxies from the DESI Bright Galaxy Survey (BGS\cite{Hahn2023}). In total, 2,473,945 galaxies with reliable spectroscopic redshifts lie within the FASHI footprint, spanning redshift ranges of $0 < z < 0.09$ and $0.25 < z < 0.41$ (see \hyperref[sec:Methods]{Methods}). Stellar masses ($M_\ast$) and star-formation rates are estimated for each galaxy via spectral energy distribution fitting. Galaxies are stacked in bins of stellar mass to derive the average \hi gas fraction, $\langle f_\hj(M_\ast)\rangle\equiv \langle M_\hj/M_\ast\rangle$.  Owing to radio frequency interference (RFI), the analysis focuses on four redshift intervals with mean redshifts $\langle z\rangle = 0.033$, $0.069$, $0.281$, and $0.358$ (see Extended Data Figure~\ref{fig:hist_z}). Corrections for optical selection effects and FAST beam confusion are described in the \hyperref[sec:Methods]{Methods}. 

\begin{figure}[t]
	\centerline{\includegraphics[width=0.7\textwidth]{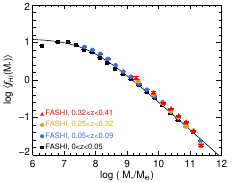}}
	\caption{\textbf{Atomic gas fraction as a function of stellar mass.} The stacked measurements of $f_\hj(M_\ast)$ in different redshift intervals are shown as filled symbols of different colours. The solid line is a best-fitting double power relation  for the measurements at $0<z<0.05$, $f_\hj(M_\ast)=12.65/[1+(M_\ast/10^{7.73}\msun)^{0.74}]$ (see other best-fitting results in Extended data Table~\ref{tab:fhifit} and Figure~\ref{fig:hismfit}).}
	\label{fig:hism}
\end{figure}

In Fig.~\ref{fig:hism} we present the stacked relation between $\langle f_\hj(M_\ast)\rangle$ and stellar mass for four redshift intervals (see also Extended Data Tables~\ref{tab:mass_z0} and~\ref{tab:mass_z3}). At low redshift ($z < 0.1$), the sample spans a broad stellar mass range, whereas at higher redshifts ($z > 0.25$) the measurements are limited to $10^9\msun$--$10^{11.5}\msun$ by the flux limits of BGS. Across all redshifts, $\langle f_\hj\rangle$ decreases systematically with increasing stellar mass and is well described by a broken power-law form, rather than a single power-law ``\hi main sequence'' reported in previous work at $M_\ast>10^9\msun$\cite{Janowiecki2020,Guo2021}. A similar double–power-law behavior is also recovered using the DESI DR1 galaxy sample \cite{Scholte2024}.

Despite this strong mass dependence, the relation exhibits remarkably little evolution with redshift. From $z = 0.41$ to the present day, $\langle f_\hj\rangle$ declines by less than $0.2$~dex at fixed stellar mass across the full range probed. This near invariance implies that the distribution of \hi gas among galaxies of different masses evolves in a highly uniform manner. Even the most massive systems ($M_\ast>10^{11}\msun$) maintain persistently low \hi gas fractions since $z\sim0.4$, consistent with their sustained quiescent star-formation activities. It indicates that the weak evolution of \hi gas fractions does not arise from changes in a particular galaxy population, but reflects a broadly similar behavior across the galaxy population as a whole.

\subsubsection{Comparison with other measurements and models}
\begin{figure}[t]
	\centerline{\includegraphics[width=\textwidth]{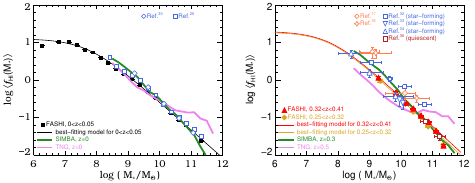}}
	\caption{\textbf{Comparisons with literature.} Left: we compare our measurements of $f_\hj$ at $0<z<0.05$ with those of Refs.\cite{Brown2015,Guo2021} (blue symbols) using ALFALFA in the same redshift range. The model predictions from the hydrodynamical simulations of TNG and SIMBA are also shown as lines of different colours. Right: our measurements of $f_\hj$ at $0.25<z<0.41$ are compared with previous measurements at similar redshifts of Refs.\cite{Rhee2018,Bera2023,Bianchetti2025,DePalma2025,Luber2025,Bianchetti2026}. We note that the measurements of Refs.\cite{Bera2023,Bianchetti2025,Luber2025} are made using only the star-forming galaxies, while that of Ref.\cite{Bianchetti2026} is for quiescent galaxies. }
	\label{fig:hism_model}
\end{figure}

In Fig.~\ref{fig:hism_model}, we compare the stellar mass dependence of the average \hi gas fraction with previous observational measurements and hydrodynamical simulations of IllustrisTNG\cite{Pillepich2018} and SIMBA\cite{Dave2019}. At $z\sim0$ (left panel), our results are consistent with earlier stacking analyses based on Arecibo data\cite{Brown2015,Guo2021}, while extending the dynamic range in stellar mass by nearly two orders of magnitude down to $10^6\msun$, as was also done previously for a DESI-selected sample\cite{Scholte2024}. Over the mass range  $10^{8.5}\msun<M_\ast<10^{10}\msun$, both TNG and SIMBA broadly reproduce the observed trends. At lower stellar masses, direct comparisons are limited by simulation resolution, whereas at the massive end ($M_\ast>10^{10.5}\msun$) TNG systematically overpredicts $f_\hj$, reflecting differences in the implementation of kinetic AGN feedback and the redistribution of atomic gas within massive halos\cite{Ma2022}.

The comparison at intermediate redshift ($z\sim0.3$; right panel in Fig.~\ref{fig:hism_model}) highlights the strength of the FASHI-DESI sample. Previous measurements of $\langle f_\hj(M_\ast)\rangle$ at these redshifts show substantial scatter, driven primarily by heterogeneous sample selections. Several studies focused exclusively on star-forming galaxies\cite{Bianchetti2025,Luber2025,Bera2023}, while others combined star-forming and quiescent populations with complex selection functions\cite{Rhee2018,DePalma2025}, complicating direct comparisons. By contrast, the DESI BGS provides a simple and well-defined selection based on r-band flux and redshift, enabling a statistically representative measurement across galaxy populations. Our measurements reconcile earlier discrepancies: at $M_\ast<10^{10}\msun$, where star-forming galaxies dominate, we recover $f_\hj$ values consistent with star-forming samples, whereas at the massive end our results agree with measurements targeting quiescent systems\cite{Bianchetti2026}. Among the simulations, SIMBA reproduces the observed $f_\hj$ in massive galaxies but overestimates the gas fractions of low-mass systems. 

\subsubsection{Cosmic HI abundance}
\begin{figure*}[t]
	\centerline{\includegraphics[width=\textwidth]{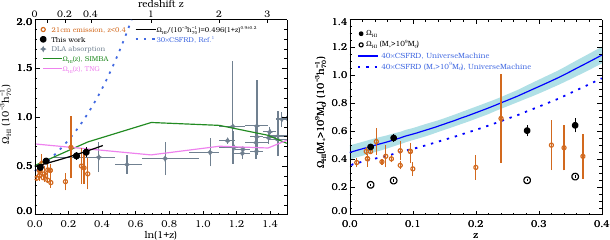}}
	\caption{\textbf{Cosmic HI gas density ($\Omega_{\rm HI}$) as a function of redshift.} Left: we compare our measurements of $\Omega_\hj$ (black filled circles) with previous results of 21-cm observations\cite{Zwaan2005,Xi2021,Guo2023,Ponomareva2023,Ma2025,Lah2007,Delhaize2013,Rhee2013,Rhee2016,Rhee2018,Bera2019,Hu2019,Chen2021,Rhee2023} (brown open circles) and DLA absorption measurements\cite{Peroux2020} (gray stars; without contribution from helium). All measurements are corrected to the same cosmological parameters. Our measurement errors of $\Omega_\hj$ show 1$\sigma$ statistical uncertainties propagated from the stacked $f_\hj$ measurements and the stellar-mass function. Forward-modelled systematic offsets are discussed in \hyperref[sec:Methods]{Methods}. Our best-fitting relation is displayed as the black solid line, $\Omega_\hj/(10^{-3}\,h_{70}^{-1})=0.496(1+z)^{0.9\pm0.2}$. The predictions from hydrodynamical simulations of TNG\cite{VillaescusaNavarro2018} and SIMBA\cite{Wen2025} are shown as the pink and green solid lines, respectively. Our measurements indicate a weak decrease of 1.35 times in $\Omega_\hj$ from $z=0.4$ to $z=0$. The corresponding decrease of the CSFRD (in units of $\msun {\rm yr}^{-1}{\rm Mpc}^{-3}$) is shown as the blue dotted line, estimated from Equation~(15) of Ref.\cite{Madau2014}. The CSFRD curve is rescaled by a constant factor of 30 for visual comparison only. Right: we show the cosmic \hi density in galaxies of $M_\ast>10^9\msun$ (open circles), where we have robust measurements of $f_\hj(M_\ast)$ in all redshift bins. The total $\Omega_\hj$ (filled circles) is shown for comparison. We also include the CSFRD estimates from the UniverseMachine model\cite{Behroozi2019} (blue solid line; times by a factor of 40) with the shaded area showing the associated errors, as well as the CSFRD contributed by galaxies with $M_\ast>10^9\msun$ (blue dotted line). More than 80\% of the CSFRD is contributed by galaxies with $M_\ast>10^9\msun$, while their $\Omega_\hj$ density is only marginally evolving. }
	\label{fig:omegahi}
\end{figure*}

Using the observed $\langle f_\hj(M_\ast)\rangle$ relations, we infer $\Omega_\hj$ by integrating over the galaxy stellar mass function (see \hyperref[sec:Methods]{Methods}). The resulting measurements are shown as black filled circles in the left panel of Fig.~\ref{fig:omegahi}. A power-law fit to the four data points yields $\Omega_\hj/10^{-3} h_{70}^{-1}=(0.496\pm0.013)(1+z)^{0.9\pm0.2}$. This fit is intended only to summarize the FASHI-DESI measurements over $0<z<0.41$, and is not extrapolated to the high-redshift regime. We find that $\Omega_\hj$ evolves only weakly over the past 4.5 Gyr, decreasing by a factor of $1.35\pm0.10$ from $z = 0.41$ to the present. Our forward-modelling mock provides estimates of the dominant systematic offsets (see \hyperref[sec:Methods]{Methods}). The two high-redshift measurements carry larger systematic uncertainties than the low-redshift measurements, mainly because of residual beam confusion and optical-selection effects. After applying these conservative offsets, the inferred decrease in $\Omega_\hj$ becomes a factor of $1.12\pm0.10$.

Compared to previous 21-cm emission measurements\cite{Rhee2016,Rhee2018,Bera2019}, our $\Omega_\hj$ constraints are systematically higher and up to three times more precise at $z > 0.25$. This improvement arises from the combination of a statistically representative galaxy sample, a well-defined optical selection function, and sensitivity to gas-rich but optically faint systems that are often under-represented in earlier studies. In Fig.~\ref{fig:omegahi} we also compare our results with the DLA measurements at higher redshifts\cite{Peroux2020} and theoretical predictions from TNG and SIMBA. While TNG predicts an approximately constant $\Omega_\hj$ over $0 < z < 0.4$\cite{VillaescusaNavarro2018}, SIMBA exhibits a modest decline by a factor of $\sim1.5$, in closer agreement with our observations\cite{Wen2025}.

The contrast between the weak evolution of $\Omega_\hj$ and the much steeper decline of CSFRD is shown in Fig.~\ref{fig:omegahi}. For reference, the CSFRD is scaled by a constant factor of 30, following the parameterization of Ref.\cite{Madau2014}, and evolves approximately as $(1+z)^{2.656}$ over the same redshift range. In the right panel of Fig.~\ref{fig:omegahi}, we further isolate the contribution from galaxies with $M_\ast>10^9\msun$, which account for more than 80\% of the CSFRD (as estimated in the empirical UniverseMachine model\cite{Behroozi2019}) but show similarly weak evolution in $\Omega_\hj$. 

\subsubsection{Implications for galaxy formation models}
Our measurements place a high-precision constraint on the late-time evolution of the cosmic baryon cycle. Over the past 4.5 Gyr, $\Omega_\hj$ decreases by only a factor of $1.35\pm0.10$ (or by a factor of $1.12\pm0.10$ after applying the conservative systematic offset; see \hyperref[sec:Methods]{Methods}), whereas the CSFRD declines by a factor of 2.46. This mild evolution in $\Omega_\hj$ is broadly consistent with the picture established by previous DLA and 21-cm measurements, but the FASHI--DESI sample provides a much tighter 21-cm emission constraint over $0<z<0.41$. As discussed in Refs.\cite{Peroux2020,Tacconi2020,Walter2020}, the late-time decline of the CSFRD is closely linked to the decline of the molecular gas reservoir, which in turn is connected to the reduced net accretion rate from the IGM and CGM as halo growth slows. Our result shows that this decline is not accompanied by a comparably rapid reduction of the \hi reservoir.

The more distinctive finding of our work is that the average \hi gas fraction $f_\hj(M_\ast)$ evolves by less than 0.2 dex over the full stellar-mass range probed, with little change in the shape of the relation. This is a substantially stronger constraint than a global $\Omega_\hj$ measurement alone. Thus, the slow evolution of $\Omega_\hj$ is not simply an integral result produced by compensating changes among different galaxy populations. Instead, it indicates that the weak evolution of \hi is a population-wide trend. The result implies that the decline in cosmological gas supply does not produce a strong mass-dependent depletion of \hi, while it has a much larger impact on the conversion of cold gas into the molecular and star-forming phases.

In a gas-regulator framework\cite{Lilly2013}, the evolution of the galaxy \hi content is governed by the balance among gas accretion from the IGM and CGM, recycling within galaxies and halos, conversion into H$_2$, and removal through ionization, outflows or environmental processes. The observed weak evolution of both $\Omega_\hj$ and $f_\hj(M_\ast)$ indicates that the late-time \hi reservoir remains close to this regulated balance, even as the throughput into molecular gas and star formation decreases substantially. A declining gas accretion rate can therefore be the upstream driver of late-time galaxy evolution, but its impact is mediated through the baryon cycle rather than appearing as primarily strong depletion of \hi itself. Possible channels include lower disk pressure, reduced shielding, changes in metallicity or turbulence, and feedback-regulated recycling. In massive halos, virial shocks can also heat accreted gas to the halo virial temperature, producing a hot and diffuse CGM with prolonged cooling times\cite{Keres2005,Dekel2006,Keres2009}. Together with AGN and other feedback processes, this can reduce the supply of cold gas to galactic disks and further suppress the formation of molecular gas and stars at late times. Our measurements therefore provide a stringent benchmark for galaxy-formation models: successful models must reproduce a baryon cycle in which $\Omega_\hj$ and $f_\hj(M_\ast)$ remain nearly invariant, while the molecular and star-forming components evolve much more strongly.

\renewcommand{\figurename}{Extended Data Figure} 
\renewcommand{\tablename}{Extended Data Table}
\section{Methods}
\label{sec:Methods}
\subsection{Cosmological parameters}
Throughout the paper, we adopt a flat $\Lambda$ cold dark matter ($\Lambda$CDM) cosmology of $\Omega_{\rm m}=0.3$, $\Omega_{\Lambda}=0.7$ and $H_0=70\,h_{70}\kms {\rm Mpc}^{-1}$.

\subsection{FASHI Observations and data}
\begin{figure*}
	\centerline{\includegraphics[width=\textwidth]{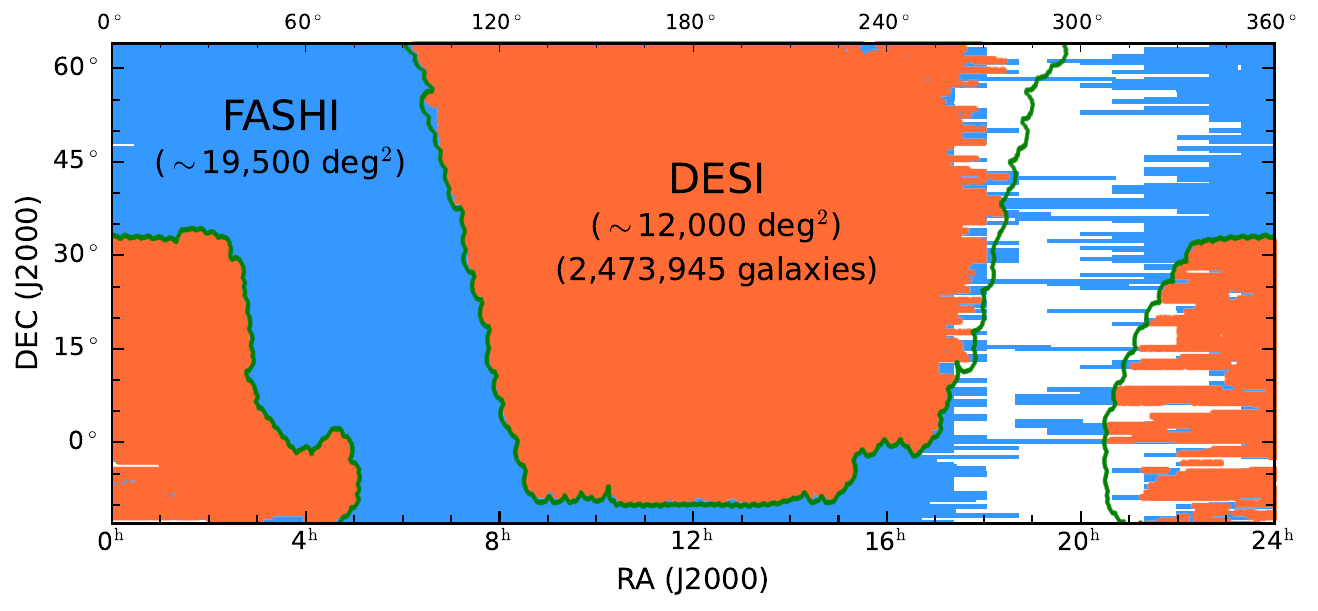}}
	\caption{\textbf{Coverage of the cross-matched FASHI and DESI sample.} FASHI has a total coverage of $\sim$19,500\,deg$^{2}$, shown in blue. DESI coverage is indicated with green curves. The cross-matched area is $\sim$12,000\,deg$^{2}$ shown in orange, yielding $2,473,945$ sources (including the confusing sources) for effective stacking in the redshift range of $0<z<0.41$.}
	\label{fig:observed_sky}
\end{figure*}
The Five-hundred-meter Aperture Spherical radio Telescope (FAST), located in Guizhou, China, has a 500~m primary aperture and an effective collecting area corresponding to a diameter of approximately 300~m, yielding a beam size of $\sim2.9$ arcmin at 1.4 GHz. The 21~cm \hi spectral data used in this work are drawn from the FAST All Sky \hi survey (FASHI\cite{Zhang2024,Zhang2026}), which employs FAST’s 19-beam receiver to scan the sky over $-14^\circ < \text{DEC} < 64^\circ$.

The spectral line backend provides a 500 MHz bandwidth (1000--1500 MHz) with 64k frequency channels, corresponding to a raw spectral resolution of 7.63 kHz (1.6\,$\kms$ at 1.4\,GHz). The pointing accuracy is maintained within $15''$. Intensity calibration was achieved by injecting a 1-second high-noise signal ($\sim$11\,K) every 32 or 64 seconds during observations. The Degrees Per Flux Unit (DPFU) for the 19 beams ranges from 13 to 17\,K\,Jy$^{-1}$ at around 1.4\,GHz \cite{Nan2011,Jiang2019,Jiang2020}.

FASHI observations were conducted primarily in drift-scan mode. The telescope's azimuth arm was positioned on the meridian at predefined J2000 declinations, spaced $21\dotmin65$ apart. The feed array was rotated by $23\dotdeg4$ to enable super-Nyquist sampling of individual beam drift tracks in declination, ensuring sampling rates below half the full width at half maximum (FWHM).

The FASHI data were processed using the \texttt{HiFAST} pipeline \cite{Jing2024}. To improve the signal-to-noise ratio, the spectral resolution was smoothed to $\sim$6.4\,$\kms$ per channel for $0<z<0.09$ and $\sim$20.0\,$\kms$ per channel for $0.25<z<0.41$. The 3D data cubes were gridded to a pixel scale of 1\,arcmin. The typical rms noise per spectrum at $0.25<z<0.41$ is $\sim$1.8\,mJy. Spectral data with rms exceeding 2.5\,mJy were excluded to reduce noise contamination. Due to severe radio frequency interference (RFI) at $0.25<z<0.41$ \cite{Zhang2022}, emission or absorption line profiles exceeding $2.0\sigma$ were masked with NaN in each spectrum.

The FASHI survey covers over 19,500\,$\deg^2$ of the sky, while the DESI survey spans approximately 15,000\,$\deg^2$. The overlapping area between FASHI and DESI is about 12,000\,$\deg^2$. The FASHI and DESI footprints are shown in Extended Data Figure\,\ref{fig:observed_sky}.

\subsection{Optical Sample of DESI galaxies}

The Dark Energy Spectroscopic Instrument (DESI) is a robotic, fibre-fed, and highly multiplexed spectroscopic survey instrument installed on the Mayall 4-meter telescope at Kitt Peak National Observatory \cite{DESI2022.KP1.Instr}. Capable of simultaneously capturing spectra for nearly 5000 objects across a $\sim3^\circ$ field of view \cite{Corrector.Miller.2023, FiberSystem.Poppett.2024}, DESI is currently conducting an eight-year survey that will cover approximately 17,000 deg$^2$ of the sky. The full survey is projected to yield about 63 million spectroscopically confirmed galaxies and quasars \cite{DESI2016b.Instr}. To manage its vast scale, the DESI project is based on a comprehensive suite of supporting software pipelines and data products \cite{Spectro.Pipeline.Guy.2023, SurveyOps.Schlafly.2023}. 
The public First Data Release (DR1) \cite{DESI2024.I.DR1} contains spectra for more than 18 million unique targets, and initial cosmological results of a full-shape analysis have been derived from this dataset \cite{DESI2024.II.KP3, DESI2024.VII.KP7B}. 
The \hi stacking analysis presented here uses an external catalogue of optical redshifts to determine galaxy positions and the expected frequency of the \hi line. The \hi data are then extracted at these coordinates and co-added. To maximise the signal-to-noise ratio of the \hi stacking, it is essential to maximise the optical sample size. Therefore, we used the forthcoming Second Data Release (DR2)  for our analysis \cite{DESI.DR2.DR2, 2025arXiv250314738D, 2025arXiv250314739D}. 

Specifically, the galaxy sky positions and redshifts are taken from the {\tt loa} release of DESI DR2, which provides spectroscopic redshifts for over 30 million galaxies across a sky area exceeding 15,000 square degrees. The reliable redshifts are selected from targets with a positive TARGETID (TARGETID > 0) and are defined as those satisfying the following criteria: \texttt{ZCAT\_PRIMARY==True}, \texttt{OBJTYPE==`TGT'}, \texttt{SPECTYPE!=`STAR'}, and \texttt{ZWARN==0}. This selection ensures that only unique, non-sky and non-star targets with no known redshift-fitting failures are included.
The redshift data in our primary focus range of $0 < z < 0.41$ are sourced from the DESI Bright Galaxy Survey (BGS) \cite{Hahn2023}. The BGS observes two primary samples: a flux-limited bright sample complete to $r < 19.5$ mag (BGS Bright) and a fainter, colour-selected sample spanning $19.5 < r < 20.175$ mag (BGS Faint). In the photometric catalog of LS-DR9~\cite{Dey2019}, the $r$-band $5\sigma$ depth for galaxies is 23.4~mag, which is three magnitudes fainter than the spectroscopic limit of $r = 20.175$. As shown in Figure~2 of Ref.~\cite{2019ApJS..242....8Z}, the $r$-band photometric completeness reaches $\sim97\%$ at this limit. To mitigate contamination from imaging artefacts and fragmented (`shredded') galaxies in the LS-DR9 data, the BGS Bright sample employs a fibre-magnitude cut of $r_\text{fibre} < 22.9$. Here, $r_\text{fibre}$ refers to the predicted r-band magnitude within a 1.5$^{\prime\prime}$-diameter aperture. The selection of BGS Faint sample uses a colour-dependent fibre-magnitude cut to ensure a high redshift success rate. Specifically, galaxies that lie above the locus defined by $(z − W1) − 1.2(g − r) + 1.2$ are more likely to exhibit detectable $H_\alpha$ or $H_\beta$ emission lines and are therefore allowed a deeper fibre-magnitude cut of $r_\text{fibre} <21.5$. In contrast, others below this locus are limited to a shallower cut $r_\text{fibre} < 20.75$.

In addition, the nearby large galaxies are often resolved into numerous substructures, which can lead to their mis-classification as dwarf galaxies. To mitigate this, we excluded all sources overlapping with the boundaries of these large galaxies—regions defined by ellipse-fitting in the Siena Galaxy Atlas (SGA\cite{2023ApJS..269....3M})—while preserving sources identified to the large galaxies themselves.

The incompleteness in the redshift catalogue is described in Ref.\cite{Hahn2023}. The BGS Bright sample will achieve a fibre assignment efficiency exceeding 80\%. Furthermore, both the BGS Bright and Faint samples will reach redshift success rates of over 95\% across a broad range of galaxies, with no significant dependence on observing conditions.
To correct it, we define a completeness weight $c_i$ that depends on the $r$-band magnitude $r$ and celestial position $\theta$. First, we construct a photometric catalog from LS DR9 using the BGS selection cuts\cite{Hahn2023}. Because we do not impose the stricter BGS-Faint magnitude limit, this catalogue serves as the complete reference sample. For a given $\theta$ and $r$, the completeness weight is defined as $c_i = N_{\rm tot} / N_z$, where $N_z$ is the number of targets with reliable redshift measurements, and $N_{\rm tot}$ is the total number of targets in the photometric sample. This approach accounts for incompleteness arising both from the sample selection itself and from the declining redshift success rate at fainter magnitudes.

Additionally, given the presence of strong RFI in the FASHI spectral data at $0<z<0.41$ \cite{Zhang2022}, we exclusively utilise data from the following redshift intervals: (0.0, 0.09), (0.25, 0.275), (0.285, 0.3), (0.31, 0.32), (0.325, 0.34), (0.345, 0.36), (0.37, 0.38), and (0.385, 0.41). The left panel of Extended Data Figure~\ref{fig:hist_z} illustrates the distribution of stacked sources and strong RFI regions across different redshifts. 

\subsection{Calculations for $M_\ast$, $L_r$, $M_{\rm HI}$, and $f_{\rm HI}$}

Stellar masses ($M_\ast$) are estimated by fitting spectral energy distributions (SEDs) using the Code Investigating GALaxy Emission (CIGALE v2022.1 \cite{2019A&A...622A.103B}). The input data for CIGALE include optical and infrared photometry in the $g$, $r$, $z$, $W1$, and $W2$ bands, obtained from Data Release 9 of the DESI Legacy Imaging Surveys (LS-DR9\cite{Dey2019}), along with spectroscopic redshifts from the {\tt loa} release of DESI DR2\cite{2025arXiv250314738D, 2025arXiv250314739D}. 
We adopt the Bruzual \& Charlot (2003) models\cite{2003MNRAS.344.1000B}, initial mass function of Chabrier (2003)\cite{2003PASP..115..763C}, and a double exponential star formation history. The formation time of the main population spans a range from 0.1 to 13 Gyr before the observation epoch, with e-folding times that span a range from 0.1 Gyr to 20 Gyr. The burst population has formation times of 10, 50, and 100 Myr, each with a fixed e-folding time of 10 Gyr. The physical model also incorporates prescriptions for nebular emission, dust attenuation\cite{2000ApJ...533..682C}, and dust emission \cite{2014ApJ...784...83D}. We explore a parameter space defined by the colour excess for nebular lines ($\rm{E(B-V)_{lines}}$, ranging from 0 to 1.6), the ionisation parameter ($\log U$; -3.0, -2.0), and the power-law slope ($\alpha$; 1.0, 1.5, 2.0).  All parameters not explicitly listed were held at their default values. Our derived stellar masses agree with those in the GALEX--SDSS--WISE Legacy Catalogue (GSWLC) \cite{2016ApJS..227....2S} over $0<z<0.2$, with a median offset of $<0.09$\,dex. We also compare with  stellar masses released in GAMA Data Release~4 \cite{2011MNRAS.418.1587T, 2022MNRAS.513..439D}, finding the median offsets of $0.04$ and $0.03$\,dex in the ranges $0.001<z<0.09$ and $0.25<z<0.42$, respectively.

The $r$-band luminosity $L_r$ of each BGS galaxy is calculated independently from that of the stellar mass. It is measured following the same steps of Ref.\cite{Wangyirong2024}, by applying $k$-corrections to $z=0.1$ using the kcorrect v4.3 code\cite{2007AJ....133..734B}.
We also apply an evolution correction following that of Ref.\cite{Loveday2015}, with the $Q_e$ parameter of $Q_e=1.03$ (their Equation~4). We refer the readers to Refs.\cite{Wangyirong2024,Loveday2015} for more details. 

The \hi mass ($M_\hj$) is calculated as follows,
\begin{eqnarray}
\frac{M_\hj}{\msun} = \frac{2.36 \times 10^5}{1+z} \left( \frac{D_L(z)}{{\rm Mpc}} \right)^2 \left( \frac{\int S_v {\rm d}v}{{\rm Jy\, km\, s}^{-1}} \right),\label{eq:himass}
\end{eqnarray}
where $z$ is the redshift, $D_L(z)$ is the luminosity distance, and $\int S_v {\rm d}v$ is the integrated line flux density. The \hi gas fraction ($f_\hj$) is defined as the ratio of \hi mass to stellar mass:
\begin{eqnarray}
f_\hj = \frac{M_\hj}{M_{\ast}},\label{eq:HIfraction}
\end{eqnarray}
where $M_\hj$ and $M_{\ast}$ are the total \hi mass and stellar mass, respectively. The stacked \hi gas fraction, $f_{\hj}^{\text{stack}}$, is obtained via weighted averaging to improve the signal-to-noise ratio (S/N), for which the noise decreases as $1/\sqrt{N}$ (see the right panel of Extended Data Figure~\ref{fig:hist_z}):
\begin{eqnarray}
f_{\hj}^{\text{stack}} = \frac{\sum_{i=1}^N w_i c_i f_{\hj,i}}{\sum_{i=1}^N w_i c_i},\label{eq:stack}
\end{eqnarray}
where $i$ indexes each spectrum included in the stack, and $N$ is the total number of spectra in the sample. Each \hi gas fraction spectrum $f_{\hj,i}$ is assigned a weighting factor $w_i$ and a completeness weight $c_i$. A common choice for the weighting scheme is $w_i = \sigma_{i}^{-2}$, with $\sigma_{\mathrm{rms},i}$ representing the noise level of the $i$th spectrum \cite{Fabello2011}. The completeness weight $c_i$ accounts for the angular completeness of each BGS galaxy. The final average \hi fraction of the sample is then obtained by integrating over the spectral channels containing the stacked detection. Extended Data Figures~\ref{fig:spec_z0.05}, \ref{fig:spec_z0.1}, and \ref{fig:spec_z0.3} present stacked \hi gas fraction spectra in units of 1/channel, allowing channel-wise summation to obtain integrated values.
\begin{table}[ht]
\centering
\caption{Cosmic HI Density Measurements}
\begin{threeparttable}
\begin{tabular}{llcllrr}
\hline
$\rm{Reference}$ & $\rm{Telescope}$ & $\rm{Method}$ & $\langle z\rangle$ & $\Omega_{\rm HI}/(10^{-3}h_{70}^{-1})$ & $N_{\rm gal}$ & $\rm{Area}/(\deg^2)$ \\ 
\hline
Ref.\cite{Zwaan2005} & $\rm{Parkes}$ & $\rm{HIMF}$ & $0.011$ & $0.375\pm0.03$ & $4,309$ & $20,626$ \\
Ref.\cite{Xi2021} & $\rm{Arecibo}$ & $\rm{HIMF}$ & $0.083$ & $0.355\pm0.030$ & $247$ & $1.35$ \\
Ref.\cite{Guo2023} & $\rm{Arecibo}$ & $\rm{HIMF}$ & $0.026$ & $0.455\pm0.029$ & $24,687$ & $6,900$ \\
Ref.\cite{Ponomareva2023} & $\rm{MeerKAT}$ & $\rm{HIMF}$ & $0.042$ & $0.526^{+0.91}_{-0.95}$ & $276$ & $4.8$ \\
Ref.\cite{Ma2025} & $\rm{FAST}$ & $\rm{HIMF}$ & $0.033$ & $0.454\pm0.02$ & $41,391$ & $31,528$ \\
\hline
Ref.\cite{Lah2007} & $\rm{GMRT}$ & $\rm{HI~stacking}$ & $0.24$ & $0.69\pm0.32$ & $121$ & $0.37$ \\
Ref.\cite{Delhaize2013} & $\rm{Parkes}$ & $\rm{HI~stacking}$ & $0.028$ & $0.402^{+0.042}_{-0.084}$ & $15,093$ & $\sim2,000$ \\
Ref.\cite{Delhaize2013} & $\rm{Parkes}$ & $\rm{HI~stacking}$ & $0.096$ & $0.456^{+0.061}_{-0.084}$ & $3,277$ & $42$ \\
Ref.\cite{Rhee2013} & $\rm{WSRT}$ & $\rm{HI~stacking}$ & $0.1$ & $0.33\pm0.05$ & $59$ & $1.4$ \\
Ref.\cite{Rhee2013} & $\rm{WSRT}$ & $\rm{HI~stacking}$ & $0.2$ & $0.34\pm0.09$ & $96$ & $1.4$ \\
Ref.\cite{Rhee2016} & $\rm{GMRT}$ & $\rm{HI~stacking}$ & $0.37$ & $0.42\pm0.16$ & $474$ & $1.7$ \\
Ref.\cite{Rhee2018} & $\rm{GMRT}$ & $\rm{HI~stacking}$ & $0.32$ & $0.50\pm0.18$ & $165$ & $<2.2$ \\
Ref.\cite{Bera2019}$^\ast$ & $\rm{GMRT}$ & $\rm{HI~stacking}$ & $0.34$ & $0.481\pm0.162$ & $445$ & $<0.32$ \\
Ref.\cite{Hu2019} & $\rm{WSRT}$ & $\rm{HI~stacking}$ & $0.066$ & $0.402\pm0.026$ & $1,895$ & $7$ \\
Ref.\cite{Chen2021} & $\rm{VLA}$ & $\rm{HI~stacking}$ & $0.051$ & $0.38\pm0.04$ & $3,622$ & $60$ \\
Ref.\cite{Rhee2023} & $\rm{ASKAP}$ & $\rm{HI~stacking}$ & $0.057$ & $0.42\pm0.08$ & $1,103$ & $60$ \\
Ref.\cite{Rhee2023} & $\rm{ASKAP}$ & $\rm{HI~stacking}$ & $0.080$ & $0.46\pm0.07$ & $2,696$ & $60$ \\
\textbf{This work} & $\rm{FAST}$ & $\rm{HI~stacking}$ & $0.033$ & $0.487\pm0.024$ & $\mathbf{193,334}$ & $\mathbf{\sim12,000}$ \\
\textbf{This work} & $\rm{FAST}$ & $\rm{HI~stacking}$ & $0.069$ & $0.551\pm0.025$ & $\mathbf{356,112}$ & $\mathbf{\sim12,000}$ \\
\textbf{This work} & $\rm{FAST}$ & $\rm{HI~stacking}$ & $0.281$ & $0.604\pm0.042$ & $\mathbf{411,101}$ & $\mathbf{\sim12,000}$ \\
\textbf{This work} & $\rm{FAST}$ & $\rm{HI~stacking}$ & $0.358$ & $0.642\pm0.050$ & $\mathbf{730,597}$ & $\mathbf{\sim12,000}$ \\
\hline
\end{tabular}
\begin{tablenotes}
\small
\item $^\ast$: This stacked sample only contains blue star-forming galaxies.
\item We have converted all the measurements to be consistent with our definition of $H_0=70\,h_{70}\kms {\rm Mpc}^{-1}$. Following common practice, the contribution of helium is not included in $\Omega_\hj$. The numbers of stacked galaxies ($N_{\rm gal}$) and the sky areas are shown in the last two columns.
\end{tablenotes}
\end{threeparttable}
\label{tab:HI_measurements}
\end{table}


\subsection{HI Stacking Method}
Directly detecting individual galaxies at $z\sim0.3$ requires a sensitivity of $\sim0.05$~mJy~beam$^{-1}$, which is challenging even with FAST~\cite{Xi2024fuds}. The sensitivity of FASHI ($\sim0.5$~mJy~beam$^{-1}$) is an order of magnitude shallower, making constructing an \hi mass function at these redshifts impractical. We therefore rely on spectral stacking to measure the average \hi properties.

We perform \hi spectral line stacking using the \hi Stacking Software (HISS) \cite{Healy2019}. The stacking procedure \cite{Healy2019} consists of four core steps: (1) inputting one-dimensional (1D) 21~cm \hi spectra and galaxy redshifts; (2) converting \hi flux density to \hi gas fraction for each 1D spectrum using Equations~(\ref{eq:himass}) and (\ref{eq:HIfraction}); (3) shifting the spectra to align galaxy signals at a relative velocity of 0~$\kms$; and (4) applying noise-optimised weights followed by co-addition to produce the final stacked spectrum with Equation~(\ref{eq:stack}). The detailed data preparation process for \hi stacking is described as follows:

Initially, three-dimensional (3D) cubelets centred on each galaxy are extracted from full data cubes with spatial dimensions of at least 12.0$\times$12.0 arcmin$^2$ and velocity coverage of around 3000~$\kms$, ensuring full coverage of each source and sufficient source background; targets near grid boundaries or with low spectral weights (prone to RFI noise or incomplete coverage) are excluded. In the DESI catalogue, the expected \hi disc sizes for almost all galaxies are smaller than 1.0 arcmin\cite{Wang2016}. However, the FAST beam size at 1.4 GHz is around 2.9 arcmin. In \hi mapping observations, the observed flux distribution of a point source may extend to roughly twice the beam size for bright sources. Consequently, integrated apertures are set with an angular diameter of 6.0 arcmin for the redshift range $0<z<0.41$, ensuring the inclusion of the majority of each galaxy's flux. After extracting 1D \hi spectra within these apertures, we apply \texttt{HiFAST} baseline subtraction to remove residual fluctuations from FASHI data reduction \cite{Baek2015}. For $0.25<z<0.41$, spectral features that exceed twice the noise level of each individual spectral line are masked with NaN and spectra with RMS $>2.5$\,mJy are rejected in the baseline subtraction, while only prominent standing waves are removed for $z<0.09$. Finally, cleaned spectra are stacked by stellar mass bins (see Extended Data Tables~\ref{tab:mass_z0} and~\ref{tab:mass_z3}), optimising signal extraction through RFI mitigation and baseline correction \cite{Xu2025}.

\subsection{Confusion Correction}

The large beam size of FAST ($\sim2.9$ arcmin) means that multiple galaxies can be enclosed within a single beam area and velocity range, leading to contamination of the extracted \hi\ signal via the so-called confusion effect \cite{Jones2016}. Previous studies have addressed confusion using various methods. For example, Ref.\cite{Delhaize2013} employed simulations to statistically correct for confusion in their stacking analysis. Other approaches include statistical models of the spatial distribution and average \hi\ masses of sources \cite{Jones2016}, or assigning estimated \hi\ masses to nearby galaxies to subtract contamination \cite{Fabello2012}. In this work, we exploit the large size of the DESI BGS sample ($\sim2.5$ million galaxies) to implement a more direct and robust approach: we construct a clean sample for stacking by removing galaxies that are likely to be confused with their neighbours, rather than applying a statistical correction.

Specifically, for the low-redshift range $0<z<0.09$, we exclude any galaxy that has a neighbour within an angular separation of $4$ arcmin and a velocity difference of $<450\ \mathrm{km\ s}^{-1}$. For the high-redshift range $0.25<z<0.41$, we adopt a more conservative angular separation of $3$ arcmin and a velocity difference of $<300\ \mathrm{km\ s}^{-1}$. As shown in Extended Data Figure~\ref{fig:spec_z0.3}, this velocity interval of $\pm300\ \mathrm{km\ s}^{-1}$ is sufficient to encompass the full \hi\ emission from individual galaxies, corresponding to a redshift span of only $\Delta z=0.001$. Accurate redshift information is therefore essential for isolating the correct signal in stacking analyses. Galaxies meeting these neighbour criteria are removed entirely from the stacking sample. This selection removes approximately $33.4\%$ of the original total sample, leaving a clean subset of galaxies for which the \hi\ signal is dominated by the target source. The large initial sample size ensures that we retain sufficient statistics for high signal-to-noise ratio stacking measurements after this cleaning process.

As shown in Extended Data Figure~\ref{fig:hism_correction}, the confusion correction is significant for galaxies with $M_\ast<10^9\msun$, where the \hi\ signals of these low-mass systems are often contaminated by nearby massive galaxies. In contrast, the correction becomes minimal for $M_\ast>10^9\msun$ at all redshifts. Nevertheless, we note that our neighbour-exclusion criteria tend to preferentially select isolated central galaxies over satellites in group environments. We further quantify the residual confusion from nearby galaxies not observed in the BGS sample using a realistic mock catalogue in the following section.

The confusion correction described above is critical for accurately deriving the low-mass slope of $f_{\hj}(M_\ast)$ at $z<0.1$; without this correction, the derived $\Omega_{\hj}$ would be significantly overestimated. The agreement of our $\Omega_{\hj}$ measurement at $0<z<0.05$ (derived from $f_{\hj}(M_\ast)$) with independent determinations from the \hi\ mass function \cite{Guo2023,Ma2025} further validates the precision of our confusion correction method.

\subsection{Luminosity Bias}

In this study, we combine the BGS Bright and Faint samples to increase the S/N of the stacked signals. However, it has no strong effect on our measurements. We checked that the differences between the fiducial measurements using the combined sample and those of using the flux-limited BGS Bright sample are very minor. The BGS Faint sample contributes only to the measurements of low-mass galaxies, where the star-forming galaxies dominate. The sample selection of BGS Faint only excludes some red galaxies\cite{Hahn2023}. The differences between the median SFRs of the BGS Bright and Faint samples are less than $0.05$~dex at all redshifts. Such a small difference in SFR will only lead to a very small variation in $M_\hj$ (expected to be less than 0.03~dex\cite{Saintonge2016}). 

\begin{figure*}[t]
	\centerline{\includegraphics[width=\textwidth]{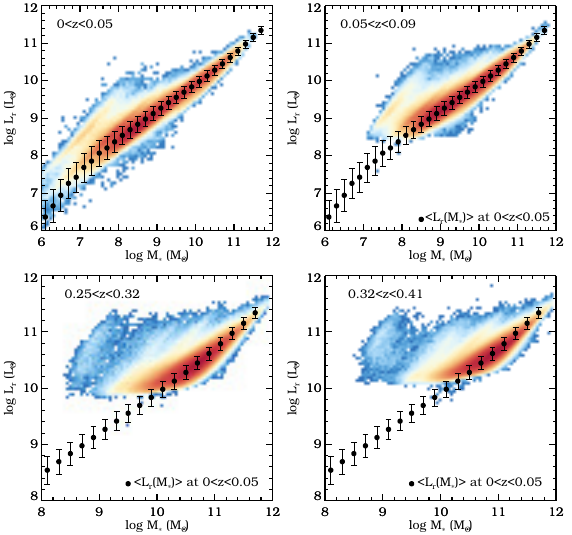}}
	\caption{\textbf{Distribution of DESI BGS galaxies as a function of stellar mass and luminosity.} The distributions of BGS galaxies in different redshift bins are shown in different panels as labelled. The colour scales represent the logarithmic number counts. The median values (and 1$\sigma$ errors) of $\langle L_r(M_\ast)\rangle$ at $0<z<0.05$ are repeated in each panel for comparisons.}\label{fig:mlum}
\end{figure*}
We note that correcting for the sample selection effect is essential to obtain the correct $f_\hj$ measurements, because the observed galaxies in flux-limited samples like BGS are biased towards more luminous (also gas-rich) galaxies\cite{Moustakas2013,Scholte2024}. As we show in Extended Data Figure~\ref{fig:mlum}, there are apparent luminosity cuts in the redshift bins of $z>0.05$. After the k-correction and the evolution correction, the $L_r$ measurements can be fairly compared at different redshift bins. The median values of $\langle L_r(M_\ast)\rangle$ at $0<z<0.05$ show good agreement with the distributions at higher redshifts for massive galaxies. It is clear that our sample is biased towards luminous galaxies at the low-mass end, especially for the two high-redshift bins. The existence of a second population of low-mass and luminous galaxies in the top left of each contour plot is caused by the same reason. These galaxies have very high SFRs compared to the overall galaxy population. The massive galaxies of $M_\ast>10^{11}\msun$ are basically complete at all redshifts. 

\begin{figure*}[t]
	\centerline{\includegraphics[width=0.8\textwidth]{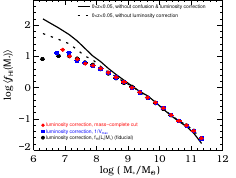}}
	\caption{\textbf{Confusion and the luminosity bias.} The original stacked measurements of $f_\hj(M_\ast)$ in $0<z<0.05$ are overestimated for the low mass end without correction for the confusion effect and luminosity bias (solid line). The measurements after the confusion correction are shown as the dotted line. The three correction methods for luminosity bias using the mass-completeness cut (red diamonds), the $1/V_{\rm max}$ weight (blue squares), and the analytical $f_\hj(L_r|M_\ast)$ (black circles) are in good agreement with each other. }\label{fig:hism_correction}
\end{figure*}
Without correction for this luminosity bias of the sample selection effect, our resulting $f_\hj$ measurements would be overestimated for low-mass galaxies, shown as the dashed line in Extended Data Figure~\ref{fig:hism_correction} for the case of $0<z<0.05$. Several methods have been proposed to correct for the luminosity selection. One is to construct a mass-complete sample for HI-stacking. Ref.\cite{Scholte2024} empirically determined the 90\% stellar mass completeness limits for BGS galaxies at $0<z<0.06$. We selected galaxies above their mass-completeness cut and measured $f_\hj(M_\ast)$ for these galaxies (shown as red diamonds). However, the drawback of this solution is the reduced number of stacked galaxies and the significantly higher stellar mass completeness limits at $z\sim0.3$ (around $10^{10.5}\msun$), which make it impossible to measure $f_\hj$ for lower-mass galaxies.  

Another typical method to correct for the flux-limited selection is to apply the $1/V_{\rm max}$ weight to each galaxy, where $V_{\rm max}$ is the maximum volume that each galaxy can be observed with the flux limit of BGS. We can simply modify Equation~(\ref{eq:stack}) by replacing $c_i$ with $c_i/V_{{\rm max},i}$ to apply this correction. The result is shown as the blue squares in Extended Data Figure~\ref{fig:hism_correction}. As expected, it is in good agreement with that of the mass-complete sample. But this method is still biased for the low-mass galaxies at $z>0.05$ where the galaxies with low $L_r$ are not observed. 

\begin{figure*}[t]
	\centerline{\includegraphics[width=\textwidth]{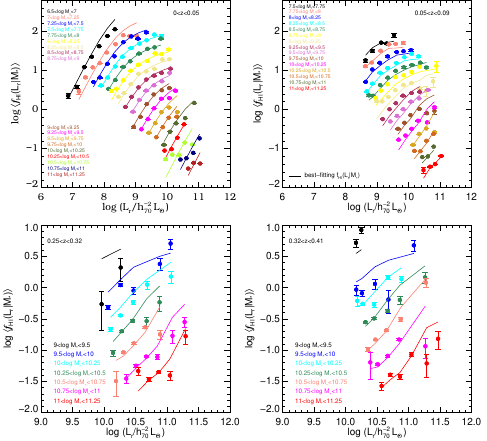}}
	\caption{\textbf{Dependence of $f_{\rm HI}(L_r|M_\ast)$ on $L_r$ in different $M_\ast$ bins.} The \hi fraction $f_\hj$ shows strong dependence on $L_r$ for a given $M_\ast$ bin at all redshifts. The measured $f_\hj(L_r|M_\ast)$ are shown as filled circles of different colours in each redshift panel. The solid lines are the best-fitting double-power model predictions of Equation~(\ref{eq:fhilm}). The shapes of $f_\hj(L_r|M_\ast)$ are all similar at different redshifts, which ensures the reliability of the $f_\hj(L_r|M_\ast)$ correction method. }\label{fig:hism_lum}
\end{figure*}
To fully exploit the BGS sample, we propose a new method to correct for the flux-limited selection. As seen in Extended Data Figure~\ref{fig:mlum}, $M_\ast$ and $L_r$ are not independent of each other. In fact, we can split the distribution of $L_r(M_\ast)$ into different $M_\ast$ and $L_r$ bins. Then we can measure the \hi fraction in each $M_\ast$ and $L_r$ bin, i.e., $f_\hj(L_r|M_\ast)$, to fully describe the dependence of $f_\hj$ on $L_r$ and $M_\ast$. Our measurements of $f_\hj(L_r|M_\ast)$ with the $1/V_{\rm max}$ corrections in the four redshift bins are shown in the different panels of Extended Data Figure~\ref{fig:hism_lum}. The $f_\hj(L_r|M_\ast)$ measurements at different redshift bins can be accurately described by the following relation, shown as solid lines.
\begin{eqnarray}
    f_\hj(L_r|M_\ast)&=&\frac{\kappa L_r}{\left(\frac{L_r}{L_\ast}\right)^{-\alpha}+\left(\frac{L_r}{L_\ast}\right)^\beta}\label{eq:fhilm}\\
    \log\kappa&=&3.347-1.345\log M_\ast+0.776\log(1+z) \\
    \log L_\ast&=&13.395-1.721\log M_\ast+0.143(\log M_\ast)^2 \\
    \alpha&=&0.596,\quad\quad \beta=0.813
\end{eqnarray}
With eight parameters, Equation~(\ref{eq:fhilm}) is able to describe the 272 data points of $f_\hj(L_r|M_\ast)$. The best-fitting model parameters are determined with the Bayesian inference tool of MULTINEST \cite{Feroz2009}. The model predictions are displayed as corresponding solid lines in each panel. At high redshift bins of $z\sim0.3$, our measurements of $f_\hj(L_r|M_\ast)$ become noisier due to the reduced number of stacked galaxies. However, as will be shown below, only the shape of $f_\hj(L_r|M_\ast)$ is relevant for the correction of the luminosity selection, which is well reproduced in different redshifts.

Then we can apply the best-fitting model of $f_\hj(L_r|M_\ast)$ with the correction factor $C_{\hj}(M_\ast)$, as follows.
\begin{eqnarray}
C_\hj(M_\ast) &=& \frac{\langle f_{\hj, {\rm intrinsic}}(M_\ast)\rangle}{\langle f_{\hj,{\rm observed}}(M_\ast)\rangle}\nonumber\\
&=& \frac{\int f_\hj(L_r|M_\ast) \phi(L_r|M_\ast) dL_r }{\int \phi(L_r|M_\ast) dL_r } \times \frac{\int N(L_r|M_\ast) dL_r}{\int f_\hj(L_r|M_\ast) N(L_r|M_\ast) dL_r }\label{eq:hicorr}
\end{eqnarray}
where $\phi(L_r|M_\ast)$ is the intrinsic distribution of $L_r$ in a given $M_\ast$ bin and $N(L_r|M_\ast)$ is the observed distribution. For simplicity, we assume no redshift dependence of $\phi(L_r|M_\ast)$ in $0<z<0.41$, which has been accounted for in the evolution correction of $L_r$ and is further supported by the comparisons shown in Extended Data Figure~\ref{fig:mlum}. We then directly used the galaxy sample at $0<z<0.05$ to infer the intrinsic distribution, i.e., $\phi(L_{r,i}|M_{\ast,i})=1/V_{{\rm max},i}$, and applied it to all other redshifts. The observed distribution $N({L_r|M_\ast})$ is measured directly from each stacked sample as $N(L_{r,i}|M_{\ast,i})=w_ic_i$, where $w_i$ and $c_i$ are the same weights as in Equation~(\ref{eq:stack}). As seen in Equation~(\ref{eq:hicorr}), only the shape of $f_\hj(M_\ast)$ matters for the correction factor $C_\hj(M_\ast)$. The result of using $C_\hj(M_\ast)$ is shown as black circles in Extended Data Figure~\ref{fig:hism_correction}, which agree well with the measurements of the other two correction methods. The measurements shown in Figure~\ref{fig:hism} are obtained by applying $C_\hj(M_\ast)$ to the original stacked $f_\hj(M_\ast)$ in each redshift bin.

We note that our evolution-correction parameter $Q_e$ is based on the measurements of Ref.\cite{Loveday2015}, which predicts an average uncertainty of $0.1$ on $Q_e$. Also, their best-fitting $Q_e$ parameters are different when separately fitting for the red ($Q_e=0.58$) and blue ($Q_e=1.09$) populations. In order to test the effect of uncertainties in $Q_e$ to our results, we construct three tests with: (1) $Q_e=1.13$, then $\log L_r$ would become $\log L_{r,{\rm new}}=\log L_r-0.04(z-0.1)$; (2) $Q_e=0.93$, then $\log L_{r,,{\rm new}}=\log L_r+0.04(z-0.1)$; (3) $Q_e=0.58$, $\log L_{r,{\rm new}}=\log L_r+0.18(z-0.1)$ for $g-r>0.15-0.03M_r^{0.1}$ and $Q_e=1.09$, $\log L_{r,{\rm new}}=\log L_r-0.024(z-0.1)$ for $g-r<0.15-0.03M_r^{0.1}$. We repeat the analysis for all three tests and find that the average differences of $\log f_\hj$ are less than $0.006$~dex for $z<0.09$ and $0.017$~dex for $0.25<z<0.41$. These errors are much smaller than other systematic uncertainties that will be investigated below, so we ignore this uncertainty in this study. This is due to the fact that the correction factor $C_\hj$ is only sensitive to the shape of $f_\hj(L_r|M_\ast)$ rather than the exact amplitude. The test (3) would potentially change the shape of $f_\hj(L_r|M_\ast)$ when applying different corrections for red and blue galaxies. Since the blue galaxies dominate the contribution to $f_\hj$, the overall correction is still very minor.

\begin{figure*}
\centerline{\includegraphics[width=0.45\textwidth]{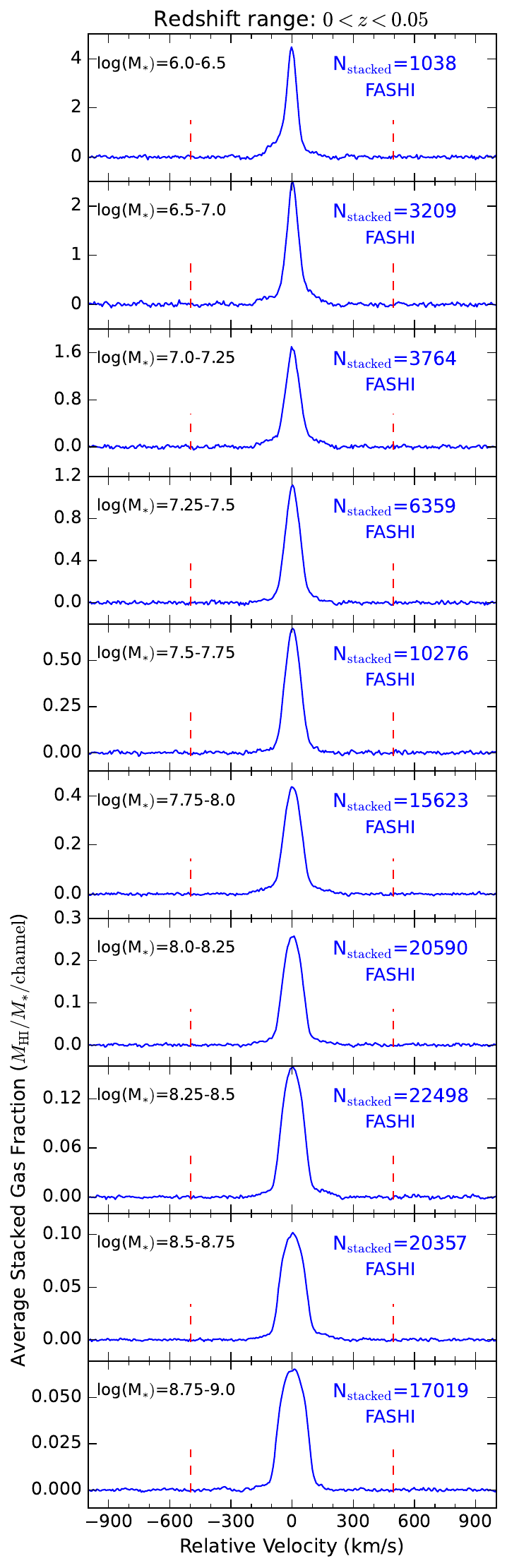}
            \includegraphics[width=0.45\textwidth]{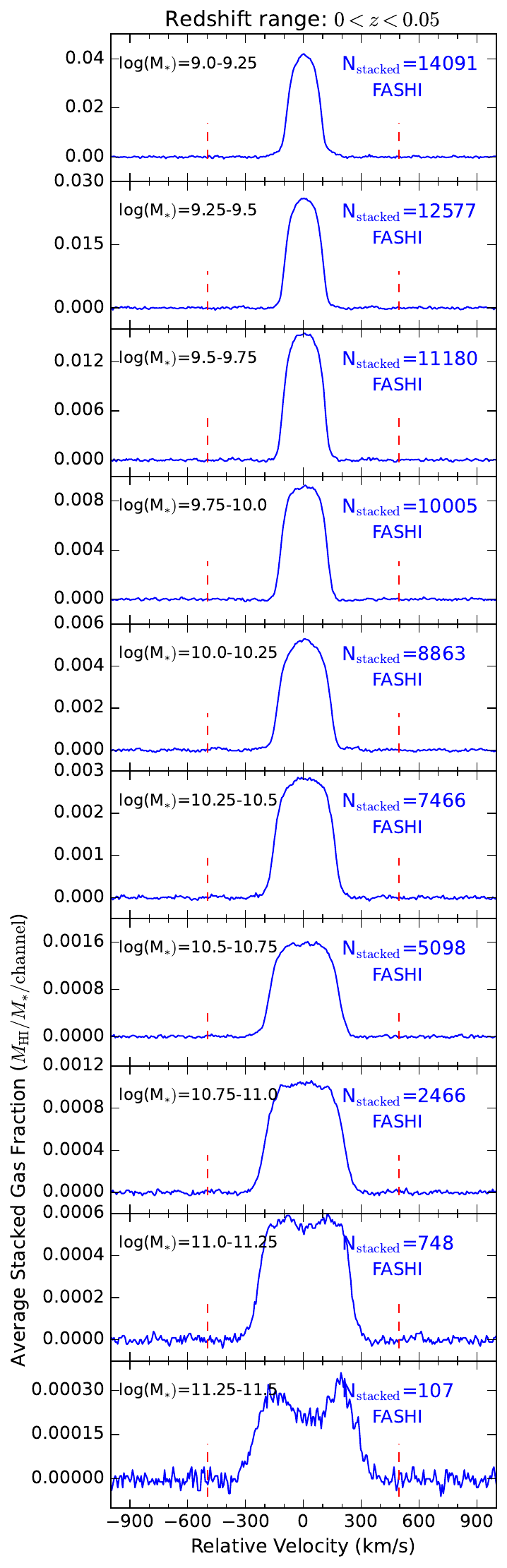}}
	\caption{\textbf{Stacked HI gas fraction at different stellar mass intervals for $0<z<0.05$.} The red vertical line shows the velocity integration limits in the $f_\hj$ calculation. $N_{\rm stacked}$ is the number of stacked spectra.} \label{fig:spec_z0.05}
\end{figure*}

\begin{figure*}
\centerline{\includegraphics[width=0.45\textwidth]{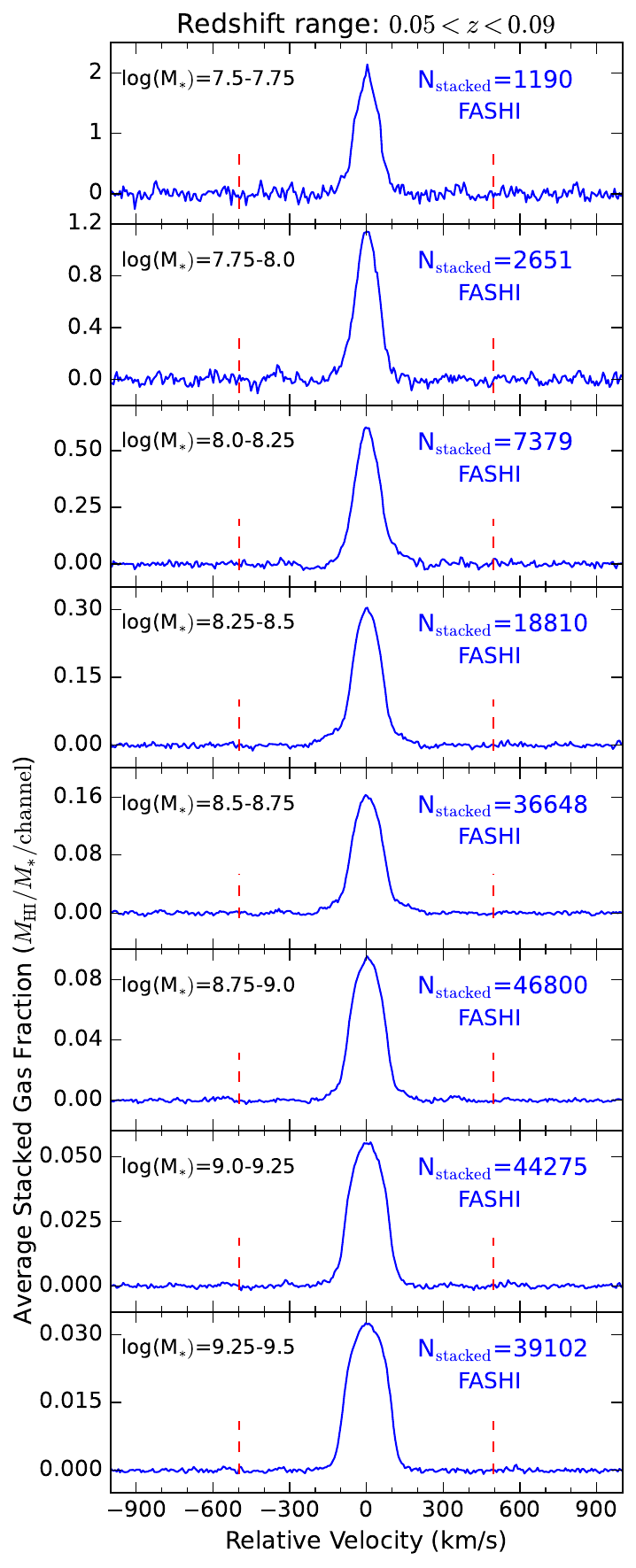}
            \includegraphics[width=0.45\textwidth]{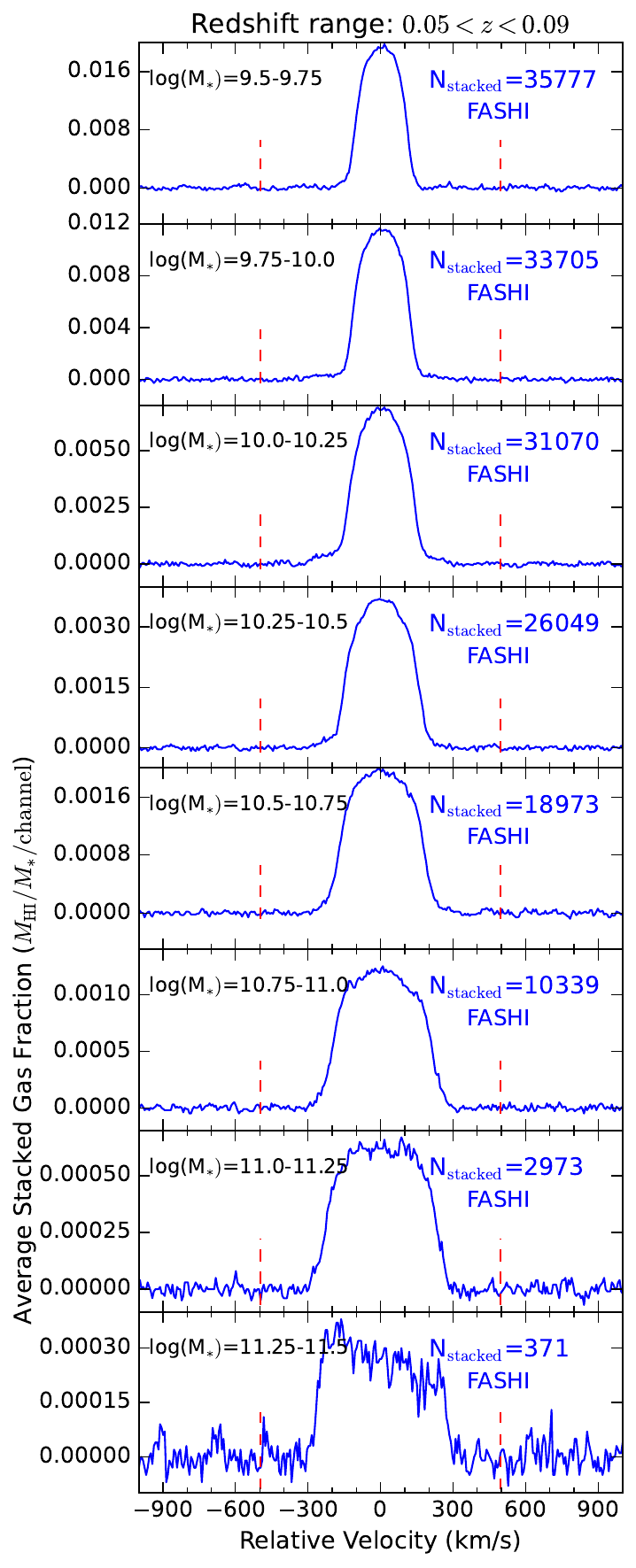}}
	\caption{\textbf{Stacked HI gas fraction at different stellar mass intervals for $0.05<z<0.09$.} The red vertical line shows the velocity integration limits in the $f_\hj$ calculation. $N_{\rm stacked}$ is the number of stacked spectra.} \label{fig:spec_z0.1}
\end{figure*}

\begin{figure*}
	\centerline{\includegraphics[width=0.45\textwidth]{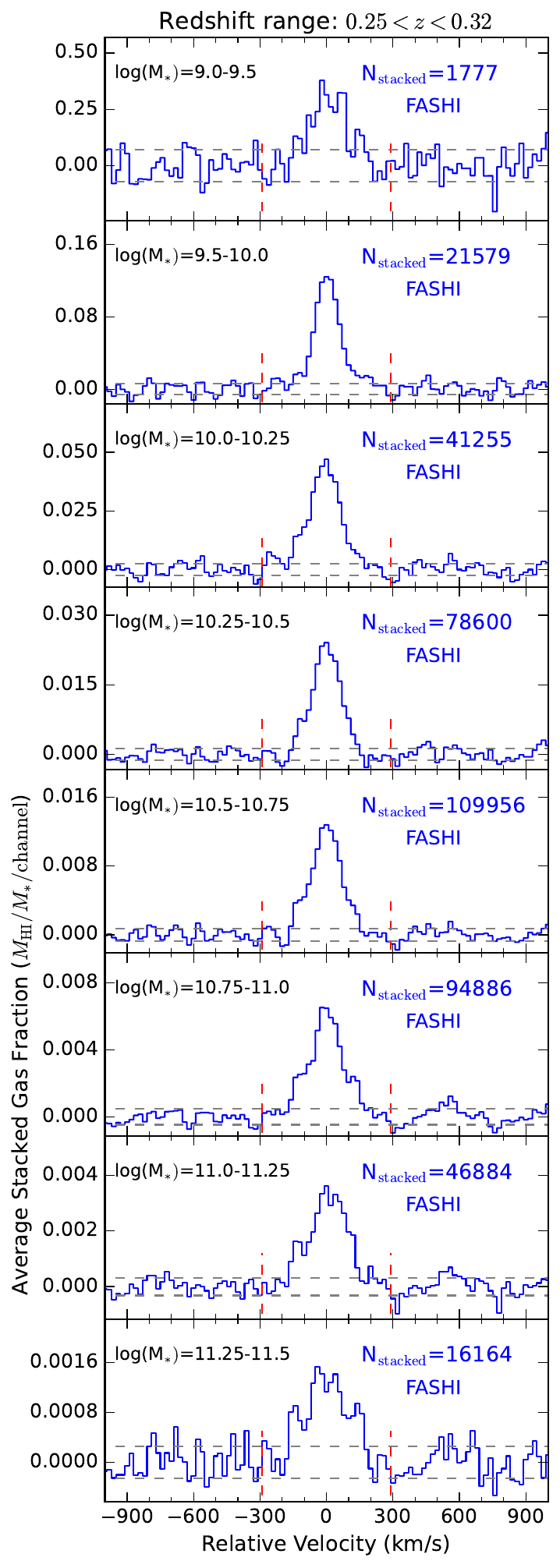}
		\includegraphics[width=0.45\textwidth]{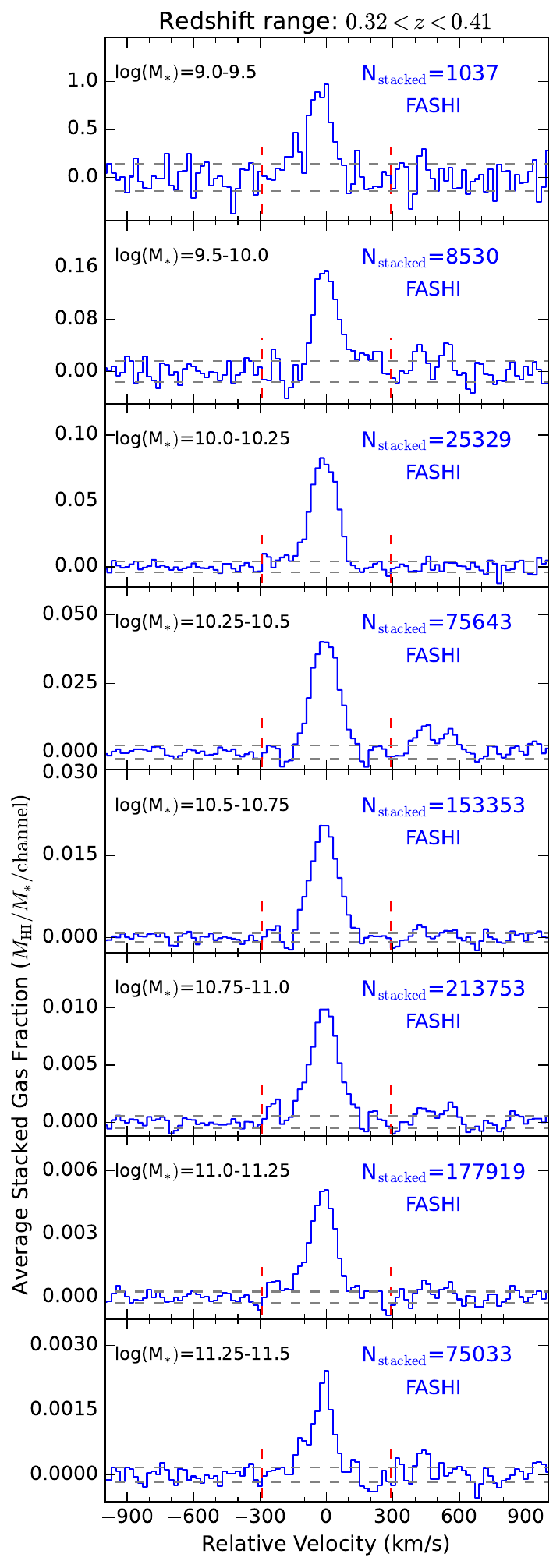}}
	\caption{\textbf{Stacked HI gas fraction at different stellar mass intervals for $0.25<z<0.32$ and $0.32<z<0.41$.} The red vertical line shows the velocity integration limits in the $f_\hj$ calculation. $N_{\rm stacked}$ is the number of stacked spectra. The gray dashed lines indicate the 1$\sigma$ noise levels. } \label{fig:spec_z0.3}
\end{figure*}

The uncertainties, expressed as the root-mean-square ($\rm rms$) values, of the stacked \hi gas fraction are estimated through a two-step process. First, we calculate the standard deviation ($\sigma$) of stacked spectra using the channels outside the integration region\cite{Sinigaglia2022,Bianchetti2025} and determine the spectral line width ($N_{\rm channel}$) from the stacked spectra presented in Extended Data Figures~\ref{fig:spec_z0.05}, \ref{fig:spec_z0.1} and \ref{fig:spec_z0.3}. $N_{\rm channel}$ is defined as the number of channels corresponding to the full width at half maximum (FWHM) of each spectral line. Subsequently, the rms uncertainties are computed using the formula ${\rm rms} = \sigma\sqrt{N_{\rm channel}}$.

\begin{figure*}
	\centerline{\includegraphics[height=0.36\textwidth]{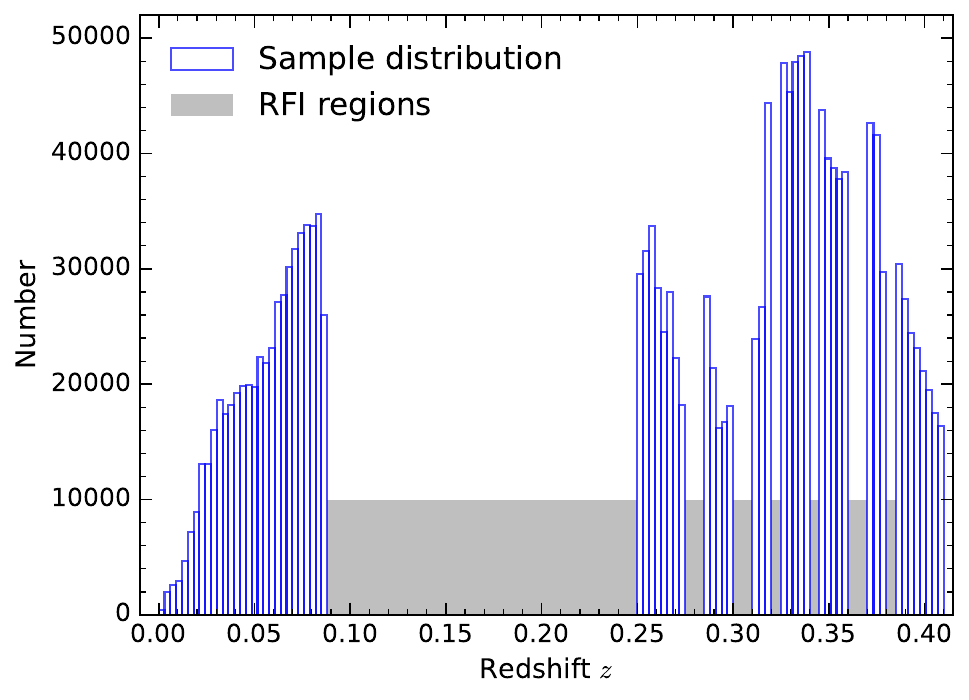}\includegraphics[height=0.36\textwidth]{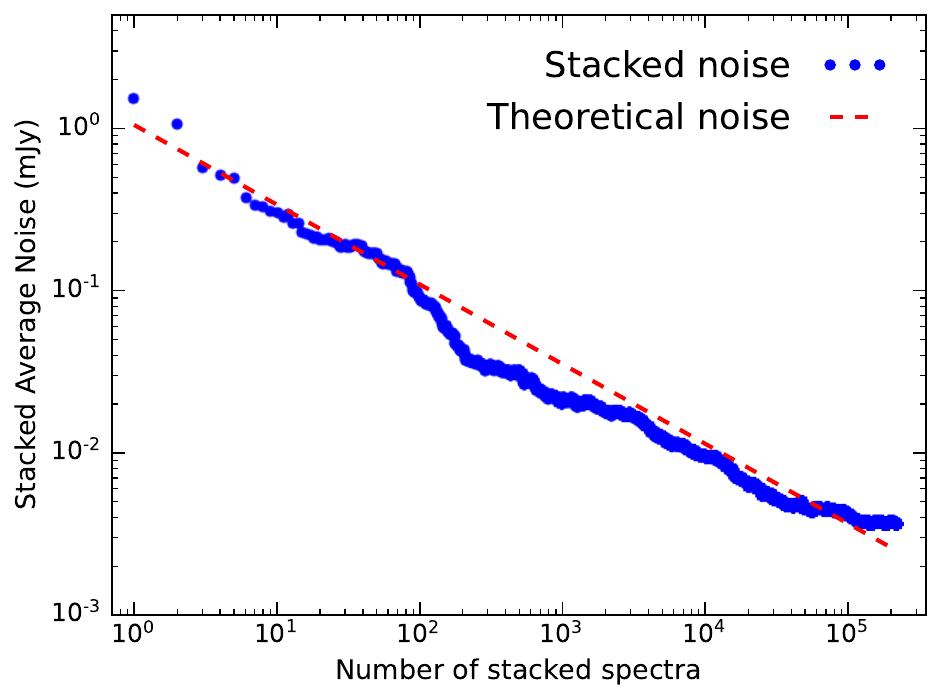}}
	\caption{\textbf{Left: distribution of stacked sources and strong RFI coverage at different redshifts.} The sample located at the strong RFI coverage has been removed before stacking. \textbf{Right: the stacked noise as a function of the number of stacked spectra for the stellar mass bin log($M_{\ast}$) = $10.75\sim11.0$ and $0.32<z<0.41$.} The red overlaid line shows the theoretical response to stacked noise: $\sigma/\sqrt{N}$, where $\sigma$ is the averaged noise for each source and $N$ is the stacked source number. }
	\label{fig:hist_z}
\end{figure*}

\subsection{Derivation of $\Omega_{\rm HI}$}
With the measurements of $f_\hj(M_\ast,z)$, we can derive the cosmic \hi abundance through the following equation,
\begin{equation}
    \Omega_\hj=\frac{1}{\rho_{\rm c}}\int \langle f_\hj(M_\ast)\rangle M_\ast\phi(M_\ast)dM_\ast, \label{eq:omega}
\end{equation}
where $\rho_{\rm c}$ is the critical density at $z=0$ and $\phi(M_\ast)$ is the galaxy stellar mass function. 

\begin{figure*}
	\centerline{\includegraphics[width=0.7\textwidth]{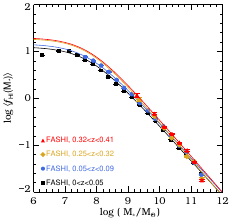}}
	\caption{\textbf{Derivation of $\Omega_{\rm HI}$.} The best-fitting models (Equation~\ref{eq:fhiz}) for $f_\hj(M_\ast)$ in different redshift bins are shown as the solid lines of different colours. The slopes at the two high redshift bins are slightly steeper.}
	\label{fig:hismfit}
\end{figure*}

\begin{table}[ht]
\centering
\caption{Best-fitting parameters for $f_\hj(M_\ast)$}
\begin{tabular}{lccc}
\hline
$z$ range & $\log f_0$ & $\log M_{\rm crit}$ & $\gamma$ \\ 
\hline
$0<z<0.05$ & $1.102\pm0.002$ & $7.730\pm0.004$ & $0.744\pm0.001$ \\
$0.05<z<0.09$ & $1.163\pm0.001$ & $7.724\pm0.015$ & $0.743\pm0.001$ \\
$0.25<z<0.32$ & $1.275\pm0.039$ & $7.730$ & $0.787\pm0.015$ \\
$0.32<z<0.41$ & $1.297\pm0.041$ & $7.730$ & $0.773\pm0.014$ \\
\hline
\end{tabular}
\label{tab:fhifit}
\end{table}
In order to accurately estimate $\Omega_\hj$ in the four redshift bins, we need to account for the limited stellar mass ranges for $z>0.05$ and the redshift evolution of $\phi(M_\ast)$. We assume that $f_\hj(M_\ast)$ has the same functional shape in the three higher redshift bins as in $0<z<0.05$. Then, we still use the following functional form to fit $f_\hj(M_\ast)$ in other redshift bins,
\begin{equation}
    f_\hj(M_\ast)=\frac{f_0}{1+(M_\ast/M_{\rm crit})^\gamma} \label{eq:fhiz}.
\end{equation}
The parameters $f_0$, $M_{\rm crit}$ and $\gamma$ describe the amplitude, characteristic mass, and slope at the massive end, respectively. For $0.25<z<0.41$, we adopted the same $\log(M_{\rm crit}/\msun)\equiv7.73$ as in $0<z<0.05$ when using Equation~(\ref{eq:fhiz}) to fit $f_\hj(M_\ast)$. This is supported by the very similar best-fitting $\log M_{\rm crit}$ ($7.724\pm0.015$) in $0.05<z<0.09$, as shown in Extended Data Table~\ref{tab:fhifit}. With the accurate measurements of $f_\hj(M_\ast)$ for $M_\ast>10^9\msun$, we are able to constrain both $f_0$ and $\gamma$ in $0.25<z<0.41$. The best-fitting curves to $f_\hj(M_\ast)$ are shown as solid lines of corresponding colours in Extended Data Figure~\ref{fig:hismfit}. The slopes $\gamma$ at the two high redshift bins are slightly steeper. 

We adopted the galaxy stellar mass function $\phi(M_\ast,z)$ from the UniverseMachine model\cite{Behroozi2019} at the corresponding redshifts. As shown in Ref.\cite{Behroozi2019}, the UniverseMachine model fits the calibrated observational measurements of $\phi(M_\ast)$ extremely well in $0<z<8$. The model prediction of $\phi(M_\ast)$ at $z<0.1$ also agrees well with the recent galaxy stellar mass function measurement of the Galaxy And Mass Assembly (GAMA) survey\cite{Driver2022}. We apply a correction of constant $M_\ast$ shift of $-0.077$~dex in $\phi(M_\ast)$ of the UniverseMachine model to match the stellar mass functions of BGS galaxies\cite{Wangyirong2024}.

Finally, we integrate Equation~(\ref{eq:omega}) for the stellar mass range of $10^6\msun<M_\ast<10^{12}\msun$ to derive the total $\Omega_\hj$. At higher redshifts, the amplitudes of $\phi(M_\ast,z)$ are slightly smaller when $f_\hj(M_\ast,z)$ becomes slightly larger, which counter-balances each other to make the evolution of $\Omega_\hj$ weaker. Unlike previous studies, our accurate measurements of $f_\hj(M_\ast)$ enable us to derive the contribution to $\Omega_\hj$ from galaxies of different stellar mass samples. The measurements of $\Omega_\hj(M_\ast>10^9\msun)$ shown in Figure~\ref{fig:omegahi} are $(0.215\pm0.007)\times10^{-3}h_{70}^{-1}$, $(0.246\pm0.008)\times10^{-3}h_{70}^{-1}$, $(0.248\pm0.008)\times10^{-3}h_{70}^{-1}$, $(0.274\pm0.010)\times10^{-3}h_{70}^{-1}$ for $0<z<0.05$, $0.05<z<0.09$, $0.25<z<0.32$, $0.32<z<0.41$, respectively. 

To reliably estimate the errors on $\Omega_\hj$, we consider both the errors of $f_\hj(M_\ast)$ and $\phi(M_\ast,z)$. The UniverseMachine model provides the errors on $\phi(M_\ast,z)$ by fitting to the observed galaxy stellar mass functions. We adopted the Monte Carlo Markov Chain (MCMC) method for fitting $f_\hj(M_\ast)$ using Equation~(\ref{eq:fhiz}) and applied the errors of $\phi(M_\ast,z)$ on each chain to derive the final errors of $\Omega_\hj$.

\subsection{Mock catalogue construction and systematic uncertainty estimate}
Along with the large statistical sample of DESI BGS galaxies, the statistical errors on $f_\hj(M_\ast)$ and $\Omega_\hj$ are significantly reduced. As a result, systematic uncertainties associated with sample selection, luminosity estimates, FAST beam confusion and sample completeness become increasingly important. To quantify the combined effects of these potential biases, we construct a forward-modelling mock catalogue based on the IllustrisTNG100-1 simulation\cite{Nelson2019}, which has sufficient resolution to represent the low-mass galaxy population probed in this work. The purpose of the mock is not to use the native \hi content of the simulation, but to use the simulated galaxy distribution, redshifts, optical properties and environments as a realistic three-dimensional tracer population for the DESI BGS observations.

The mock construction proceeds as follows.

(1) Light-cone creation:

We first construct a light-cone catalogue from TNG100-1 snapshots covering the redshift range from $z\simeq0$ to $z\simeq0.41$ (snapshots 71 to 99). Each snapshot is assigned to a radial shell, with shell boundaries defined by the midpoints in comoving distance between adjacent snapshots. The simulation cube is periodically replicated to cover the required angular and redshift range. To reduce artificial repetition, each replicated cube is randomly translated and rotated using the rotational symmetries of the cube. For each mock galaxy, we compute the sky coordinates, cosmological redshift, line-of-sight peculiar velocity and observed redshift.

(2) Stellar mass and photometry:

The mock retains the stellar mass and rest-frame optical photometry, including the $g$- and $r$-band absolute magnitudes, from the TNG catalogue. We correct the TNG $r$-band absolute magnitudes as $M_{r,{\rm corr}}=M_r+1+1.6(z-0.1)$, to match the average evolution-corrected luminosities of BGS galaxies. We also apply an average $k$-correction, $K(z)=-0.21+0.72z+1.37z^2$, derived from the BGS sample to determine the final $r$-band apparent magnitudes. The mock is then restricted to a sky area of $110^\circ<{\rm RA}<240^\circ$ and $-10^\circ<{\rm Dec}<50^\circ$ to reduce the sample size. The redshift intervals are matched to the four FASHI stacking bins, $0<z<0.05$, $0.05<z<0.09$, $0.25<z<0.32$ and $0.32<z<0.41$, with the same high-redshift RFI windows as adopted in the data analysis (Extended Data Figure~\ref{fig:hist_z}).

(3) DESI Sample selection:

We apply the DESI angular mask to the mock galaxies. The final sky area of the mock sample is $\sim6300\,{\rm deg}^2$, containing a total of 0.93 billion galaxies, which is sufficient for testing the relevant systematic effects. We then apply the BGS tiling strategy\cite{2025JCAP...01..127L,2025JCAP...01..125R} and fibre-assignment pipeline \texttt{fiberassign}\footnote{\url{https://github.com/desihub/fiberassign}} to mimic the observations. A cut in apparent magnitude of $r<19.5$ mag is imposed to select BGS Bright galaxies. For BGS Faint galaxies, we impose the flux limit $19.5 <r < 20.175$ and the fibre magnitude cut, which approximately mimics the BGS Faint selection.
The average redshift success rates for BGS Bright (99\%) and Faint galaxies (97\%) are then incorporated into the mocks.

(4) Fibre-magnitude completeness: 

To account for the fibre-magnitude selection and the associated incompleteness of extended low-surface-brightness galaxies, we estimate a mock fibre magnitude for each galaxy. The stellar half-mass radius in the TNG100 light cone is used as a proxy for the projected $r$-band half-light radius and is converted to an angular effective radius, $r_{\rm e}$, using the angular-diameter distance at the observed redshift. Assuming an exponential surface-brightness profile, the fraction of the total $r$-band flux enclosed within the 1.5 arcsec-diameter DESI fibre is
\begin{equation}
f_{\rm fib}=1-e^{-x}(1+x), \qquad
x=1.678(0.75^{\prime\prime}/r_{\rm e}), 
\end{equation}
and the corresponding mock fibre magnitude is
\begin{equation}
r_{\rm fibre}=r_{\rm app}-2.5\log f_{\rm fib}.
\end{equation}
At fixed total magnitude, galaxies with larger angular sizes have fainter fibre magnitudes and are therefore more likely to be rejected. For the BGS Faint sample, we apply the stricter fibre-magnitude requirement $r_{\rm fibre}<20.75$ in the fiducial mock as a conservative choice, to test the maximum plausible impact of this low-surface-brightness selection. This procedure propagates the uncertainty associated with the BGS fibre-magnitude selection, and in particular the possible loss of low-surface-brightness galaxies, into our final systematic error budget. The resulting galaxy number-density distributions agree well with those of the observed BGS samples. For galaxies satisfying the BGS selection, we record whether they are assigned a fibre and assign the corresponding angular completeness weight, $c_{\rm fibre}$. This allows the mock to be analysed with the same weighted estimator as used for the observed FASHI--DESI sample.

(5) \hi assignment: 

\gh{We note that a larger \hi content in massive galaxies of the TNG post-processed \hi catalogues\cite{Diemer2018} would increase the predicted confusion from massive neighbours.}
We then assign \hi masses empirically because the aim of the mock is to isolate observational systematics rather than to test the physical \hi prediction of the simulation. The initial \hi gas fraction, $f_\hj$, is assigned using our best-fitting relation $f_\hj(L_r|M_\ast)$ (Eq.~\ref{eq:fhilm}), so that the mock preserves the observed correlation between \hi content and optical luminosity at fixed stellar mass.

We next apply an environment-dependent \hi suppression to mimic the reduced \hi content of satellites and galaxies close to massive haloes due to tidal and ram-pressure stripping \cite{Zhang2013,Guo2023}. Around every central halo with virial mass $M_{200}>10^{12}\,\msun$, we search for neighbouring galaxies within $3R_{200}$. If a galaxy lies within more than one such halo, the smallest host-centric distance in units of the virial radius, $R_{200}$, is adopted. The environmental correction is applied as
\begin{eqnarray}
\Delta \log f_{\rm HI}&=&\frac{A}{2}\left[1+{\rm erf}\left(\frac{d_{\rm host}-0.8}{1.5}\right)\right]-1,\\
A&=&0.65+0.5\log\left(10^{10.2}\msun/M_\ast\right),\quad d_{\rm host}=R/R_{200},
\end{eqnarray}
which approximately reproduces the observed \hi depletion in dense environments \cite{Zhang2013}.

After applying the environmental correction, the \hi assignment is renormalized in each stellar-mass and redshift bin so that the mean $f_\hj$ of all mock galaxies reproduces the fiducial relation in Eq.~(\ref{eq:fhiz}). This step ensures that any recovered offset in the mock reflects selection and confusion biases, rather than an arbitrary difference in the input \hi normalization.

(6) Observational selections and confusion:

The mock is then passed through the same sequence of observational selections as the data. We select all BGS galaxies with assigned fibres and apply the same FAST beam-confusion removal method as in the observations. At low redshift, we reject targets that have a fibre-assigned BGS neighbour within 4 arcmin and $|\Delta v|<450\,{\rm km\,s^{-1}}$. At high redshift, the corresponding cuts are 3 arcmin and $|\Delta v|<300\,{\rm km\,s^{-1}}$.

For targets passing the clean-sample selection, we estimate the residual \hi contamination from galaxies that are not included in the BGS sample, including optically fainter galaxies that fail the BGS selection cuts and BGS targets that are not assigned fibres. The \hi contribution from faint satellite galaxies around central galaxies can contaminate the final stacked signal. To estimate this residual contamination, we follow the method described in the appendix of Ref.~\cite{Fabello2012}.

For each target, we first search for contaminating galaxies with projected distances smaller than the FAST beam FWHM, $\theta_{\rm FWHM}$, and velocity separations smaller than $300\,\kms$ in each redshift bin. We adopt a Gaussian beam response function,
\begin{equation}
B(\theta)=\exp\left[-4\ln 2\left(\frac{\theta}{\theta_{\rm FWHM}}\right)^2\right],
\end{equation}
where $\theta$ is the angular separation between the contaminating galaxy and the stacked target. Since only part of the contaminating flux contributes within the velocity window of the target, we assume a Gaussian \hi line profile for each galaxy. Unlike Ref.~\cite{Fabello2012}, which adopted a box-shaped line profile for simplicity, the Gaussian profile allows a more accurate estimate of the flux contamination. The profile FWHM is taken directly from the best-fitting $W_{50}$ of the stacked profiles in Extended Data Figures~\ref{fig:spec_z0.05}, \ref{fig:spec_z0.1}, and~\ref{fig:spec_z0.3}. We denote by $F(\Delta v)$ the fraction of the contaminating galaxy's Gaussian line profile falling within the target-centred velocity window. Following Ref.~\cite{Fabello2012}, we only include contamination from contributing galaxies with $|\Delta v|<50\,\kms$, whose \hi flux is fully blended with the target. The final contaminating contribution to the observed \hi gas fraction of target galaxy $i$ is calculated as
\begin{equation}
f_{{\rm HI},i}^{\rm obs}=f_{{\rm HI},i}+\sum_j\frac{M_{\hj,j}}{M_{*,i}}\frac{D_{L,i}^{2}/(1+z_i)}{D_{L,j}^{2}/(1+z_j)}B(\theta_{ij})F(\Delta v_{ij}),
\end{equation}
where $f_{{\rm HI},i}$ is the true target \hi fraction and $D_L$ is the luminosity distance.

(7) $f_\hj$ measurement in the mock:

We measure $f_\hj(L_r|M_\ast)$ using the BGS-selected mock galaxies weighted by $c_{\rm fibre}/V_{\rm max}$. We then fit the mock $f_\hj(L_r|M_\ast)$ relation in the same luminosity and stellar-mass bins as in the observations, using Eq.~(\ref{eq:fhilm}). We note that the fitting parameters are different from the original ones in Eq.~(\ref{eq:fhilm}). Because the luminosity correction factor $C_\hj(M_\ast)$ depends only on the relative shape of Eq.~(\ref{eq:fhilm}) in each $M_\ast$ bin, we allow the mass-bin normalization, $\kappa$, to float as a nuisance amplitude when fitting the mock $f_\hj(L_r|M_\ast)$. The factor $C_\hj(M_\ast)$ is determined in the same way as in the observations, using the mock redshift bin $0<z<0.05$ as the reference sample to infer the intrinsic luminosity distribution.

(8) Systematic error estimation:

Finally, we measure the stacked $f_\hj(M_\ast)$ for mock galaxies weighted by $c_{\rm fibre}w_{\rm noise}$, where $w_{\rm noise}$ is the rms-noise weight. Assuming a constant rms noise in flux density, the noise in $f_\hj$ is proportional to $D_L^2/[(1+z)M_\ast]$. We therefore adopt an inverse-variance weight of $w_{\rm noise}=(1+z)^2M_\ast^2/D_L^4$. The correction factor $C_\hj(M_\ast)$ is then applied to the mock-observed $f_\hj$ to recover the intrinsic $f_\hj$. The recovered mock measurement after this correction is directly compared with the input fiducial $f_{\rm HI}(M_\ast)$ relation. The remaining offset is taken as the systematic uncertainty associated with the combined effects of DESI sample selection, fibre assignment, luminosity-dependent incompleteness and residual FAST beam confusion.

\begin{figure*}
	\centerline{\includegraphics[width=1\textwidth]{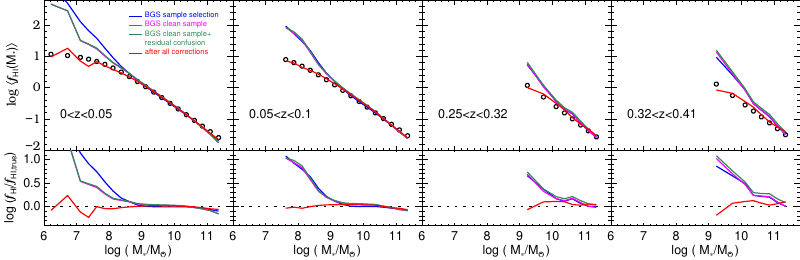}}
	\caption{\textbf{Systematic uncertainty estimates.} Top panels: we construct mock galaxy catalogues to estimate the systematic uncertainties in $f_\hj(M_\ast)$ associated with BGS sample selection (blue lines), clean-sample cuts (magenta lines), added residual confusion (green lines) and the final corrections (red lines). The black circles show the input measurements renormalized to the best-fitting models of Eq.~(\ref{eq:fhiz}). Bottom panels: offsets between the relevant $f_\hj$ measurements in the top panels and the input relation.}
	\label{fig:sys}
\end{figure*}

The effects of BGS sample selection, clean-sample cuts, residual confusion and the final corrections are shown in Extended Data Figure~\ref{fig:sys}. For $0<z<0.05$, the correction at $\log(M_\ast/\msun)<7.5$ fluctuates relatively strongly because of the small number of low-mass galaxies remaining in the mock after all selections. The average correction at $z<0.09$ is less than 0.02 dex, while the average correction at $0.25<z<0.41$ is slightly larger, around 0.05 dex, owing to the higher level of confusion in the high-redshift samples.

Since the systematic errors in $f_\hj(M_\ast)$ are correlated across stellar-mass bins, they do not map directly onto the systematic uncertainty in $\Omega_\hj$. To estimate the systematic uncertainty in our $\Omega_\hj$ measurement, we follow the same procedure as for the observations: we fit the mock $f_\hj(M_\ast)$ relation using Eq.~(\ref{eq:fhiz}) to derive the best-fitting $\Omega_\hj$ in each redshift bin. Compared with the mock truth, the systematic offsets in $\Omega_\hj$ for the redshift bins $0<z<0.05$, $0.05<z<0.09$, $0.25<z<0.32$ and $0.32<z<0.41$ are $-0.012\times10^{-3}h_{70}^{-1}$, $-0.013\times10^{-3}h_{70}^{-1}$, $+0.038\times10^{-3}h_{70}^{-1}$ and $+0.082\times10^{-3}h_{70}^{-1}$, respectively. These systematic offsets are smaller than the observed difference between the evolution of $\Omega_\hj$ and the CSFRD, and do not change our main conclusion. After applying the systematic offsets, the factor of decrease in $\Omega_\hj$ from $z=0.41$ to today becomes $1.12\pm0.10$. The majority of the systematic uncertainty arises from residual confusion.

We note that these systematic uncertainties are still conservative estimates. For example, we adopt the strict cut $r_{\rm fibre}<20.75$ for all BGS Faint galaxies, rather than the colour-dependent cuts used in the real observations. In addition, our simplified treatment of $r_{\rm fibre}$ may introduce residual confusion from some faint galaxies that could otherwise pass the BGS selection. As seen from the comparison between Extended Data Figures~\ref{fig:hism_correction} and~\ref{fig:sys}, the effects of BGS sample selection and confusion are likely overestimated in the mocks.

We also constructed a TNG50-based mock catalogue to assess the sensitivity of the forward-modelled systematic correction to numerical resolution\cite{Pillepich2019}. The TNG50 lightcone was processed through the same pipeline as the fiducial TNG100 mock. We randomly downsampled the galaxies in TNG50 to match the observed BGS number density distribution. The TNG50 mock yields a systematic trend in $f_\hj(M_\ast)$ that is consistent with the fiducial TNG100 result. \gh{This agreement suggests that the inferred systematic bias is not driven by the limited mass resolution of the TNG100 simulation.} We nevertheless use TNG100 as the fiducial mock because the dominant uncertainties in this application are related to galaxy distributions in various large-scale environments, for which the larger simulated volume is preferable. The TNG50 result provides a robustness check for the fiducial correction.

\begin{table*}
\renewcommand{\arraystretch}{1.2} 
\caption{Measurements of the HI-stellar mass relation at $0<z<0.05$ and $0.05<z<0.09$.}
\label{tab:mass_z0}
\centering
\begin{tabular}{l|rrcr|rrcr} 
 \hline
 & \multicolumn{4}{c|}{$0<z<0.05$} & \multicolumn{4}{c}{$0.05<z<0.09$}  \\
$\log(M_\ast/\msun)$ & $\langle\log M_\ast\rangle$ & $\log f_\hj$ & $\sigma_{\log f_\hj}$ & $N_{\rm gal}$ & $\langle\log M_\ast\rangle$ & $\log f_\hj$ & $\sigma_{\log f_\hj}$ & $N_{\rm gal}$ \\
 \hline
$[ 6.0 ,\, 6.5 ]$& $6.284$ & $0.944$& $0.003$& $1038$&  \\
$[ 6.5 ,\, 7.0 ]$& $6.797$ & $1.050$& $0.003$& $3209$&  \\
$[ 7.0 ,\, 7.25 ]$& $7.136$ & $1.047$& $0.003$& $3764$&  \\
$[ 7.25 ,\, 7.5 ]$& $7.387$ & $0.950$& $0.003$& $6359$&  \\
$[ 7.5 ,\, 7.75 ]$& $7.634$ & $0.822$& $0.002$& $10276$& $7.645$ & $0.883$& $0.009$& $1190$ \\
$[ 7.75 ,\, 8.0 ]$& $7.883$ & $0.768$& $0.002$& $15623$& $7.896$ & $0.803$& $0.006$& $2651$ \\
$[ 8.0 ,\, 8.25 ]$& $8.128$ & $0.621$& $0.002$& $20590$& $8.149$ & $0.700$& $0.003$& $7379$ \\
$[ 8.25 ,\, 8.5 ]$& $8.374$ & $0.478$& $0.002$& $22498$& $8.393$ & $0.566$& $0.003$& $18810$ \\
$[ 8.5 ,\, 8.75 ]$& $8.622$ & $0.330$& $0.002$& $20357$& $8.634$ & $0.395$& $0.002$& $36648$ \\
$[ 8.75 ,\, 9.0 ]$& $8.871$ &$0.174$& $0.001$& $17019$& $8.876$ & $0.222$& $0.002$& $46800$ \\
$[ 9.0 ,\, 9.25 ]$& $9.122$ & $0.029$& $0.001$& $14091$& $9.122$ & $0.087$& $0.002$& $44275$ \\
$[ 9.25 ,\, 9.5 ]$& $9.372$ & $-0.137$& $0.001$& $12577$& $9.373$ & $-0.102$& $0.002$& $39102$ \\
$[ 9.5 ,\, 9.75 ]$& $9.623$ &$-0.317$& $0.001$& $11180$& $9.625$ & $-0.259$& $0.001$& $35777$ \\
$[ 9.75 ,\, 10.0 ]$& $9.873$ & $-0.492$& $0.001$& $10005$& $9.873$ & $-0.437$& $0.001$& $33705$ \\
$[ 10.0 ,\, 10.25 ]$& $10.121$ & $-0.677$& $0.001$& $8863$& $10.123$ & $-0.598$& $0.001$& $31070$ \\
$[ 10.25 ,\, 10.5 ]$& $10.369$ & $-0.878$& $0.001$& $7466$& $10.370$ & $-0.815$& $0.001$& $26049$ \\
$[ 10.5 ,\, 10.75 ]$& $10.616$ & $-1.067$& $0.001$& $5098$& $10.617$ & $-1.008$& $0.001$& $18973$ \\
$[ 10.75 ,\, 11.0 ]$& $10.854$ & $-1.209$& $0.001$& $2466$& $10.855$ & $-1.151$& $0.001$& $10339$ \\
$[ 11.0 ,\, 11.25 ]$& $11.089$ & $-1.380$& $0.002$& $748$& $11.091$ & $-1.372$& $0.003$& $2973$ \\
$[ 11.25 ,\, 11.5 ]$& $11.328$ & $-1.627$& $0.004$& $107$& $11.327$ & $-1.629$& $0.008$& $371$ \\
\hline
\end{tabular}
\end{table*}

\begin{table*}
\renewcommand{\arraystretch}{1.2} 
\caption{Measurements of the HI-stellar mass relation at $0.25<z<0.41$.}
\label{tab:mass_z3}
\centering
\begin{tabular}{l|rrcr|rrcr} 
 \hline
 & \multicolumn{4}{c|}{$0.25<z<0.32$} & \multicolumn{4}{c}{$0.32<z<0.41$}  \\
log($M_\ast/M_\odot$)& $\langle\log M_\ast\rangle$ & $\log f_{\hj}$ & $\sigma_{\log f_\hj}$ & $N_{\rm gal}$ & $\langle\log M_\ast\rangle$ & $f_\hj$ & $\sigma_{\log f_\hj}$ & $N_{\rm gal}$ \\
 \hline
$[ 9.0 ,\, 9.5 ]$& $9.347$ & $-0.025$& $0.047$& $1777$& $9.286$ & $0.066$& $0.049$& $1037$ \\
$[ 9.5 ,\, 10.0 ]$& $9.843$ & $-0.378$& $0.016$& $21579$& $9.842$ & $-0.329$& $0.022$& $8530$ \\
$[ 10.0 ,\, 10.25 ]$& $10.141$ & $-0.634$& $0.015$& $41255$& $10.154$ & $-0.605$& $0.014$& $25329$ \\
$[ 10.25 ,\, 10.5 ]$& $10.386$ & $-0.842$& $0.013$& $78600$& $10.394$ & $-0.761$& $0.017$& $75643$ \\
$[ 10.5 ,\, 10.75 ]$& $10.628$ & $-0.988$& $0.014$& $109956$& $10.636$ &$-0.946$& $0.011$& $153353$ \\
$[ 10.75 ,\, 11.0 ]$& $10.865$ & $-1.241$& $0.021$& $94886$& $10.878$ &$-1.110$& $0.015$& $213753$ \\
$[ 11.0 ,\, 11.25 ]$& $11.106$ & $-1.352$& $0.019$& $46884$& $11.113$ &$-1.350$& $0.016$& $177919$ \\
$[ 11.25 ,\, 11.5 ]$& $11.348$ & $-1.727$& $0.042$& $16164$& $11.348$ & $-1.746$& $0.027$& $75033$ \\
\hline
\end{tabular}
\end{table*}

{\footnotesize
\bibliographystyle{naturemag}
@preamble{"\providecommand{\noopsort}[1]{}"}

}

\section{Acknowledgments}
We thank Luis C. Ho, Cheng Cheng, Jie Wang, Jing Wang, Jin-Long Xu, Nai-Ping Yu, and Xiao-Lan Liu for helpful discussions and contributions to the FASHI data processing. This work is supported by the National Natural Science Foundation of China (Nos.\,12225303, 12288102, 12421003, 12595313, 12373001, 12120101, 12503013, 12373010), the National Key R\&D Program of China (Nos.\,2025YFE0202300, 2022YFA1602902, 2023YFA1607804), the National SKA Program of China (No.\,2025SKA0150100), the science research grants from the China Manned Space Project (No.\,CMS-CSST-2021-A05), the Guizhou Provincial Science and Technology Projects (Nos.\,QKHFQ[2023]003, QKHPTRC-ZDSYS[2023]003, QKHFQ[2024]001, QKHJCMS[2025]015), the CAS Project for Young Scientists in Basic Research (Nos.\,YSBR-063, YSBR-092), and the Office of Science and Technology, Shanghai Municipal Government (Nos.\,24DX1400100, ZJ2023-ZD-001). FAST is a Chinese national mega-science facility, operated by the National Astronomical Observatories of Chinese Academy of Sciences (NAOC).  

This research used data obtained with the Dark Energy Spectroscopic Instrument (DESI). DESI construction and operations is managed by the Lawrence Berkeley National Laboratory. This material is based upon work supported by the U.S. Department of Energy, Office of Science, Office of High-Energy Physics, under Contract No.\,DE–AC02–05CH11231, and by the National Energy Research Scientific Computing Center, a DOE Office of Science User Facility under the same contract. Additional support for DESI was provided by the U.S. National Science Foundation (NSF), Division of Astronomical Sciences under Contract No.\,AST-0950945 to the NSF’s National Optical-Infrared Astronomy Research Laboratory; the Science and Technology Facilities Council of the United Kingdom; the Gordon and Betty Moore Foundation; the Heising-Simons Foundation; the French Alternative Energies and Atomic Energy Commission (CEA); the National Council of Humanities, Science and Technology of Mexico (CONAHCYT); the Ministry of Science and Innovation of Spain (MICINN), and by the DESI Member Institutions: www.desi.lbl.gov/collaborating-institutions. The DESI collaboration is honored to be permitted to conduct scientific research on I’oligam Du’ag (Kitt Peak), a mountain with particular significance to the Tohono O’odham Nation. Any opinions, findings, and conclusions or recommendations expressed in this material are those of the author(s) and do not necessarily reflect the views of the U.S. National Science Foundation, the U.S. Department of Energy, or any of the listed funding agencies. 

\section{Author Contributions}
C.Z., H.G. and Y.G. contributed equally to this work, and H.G. and Y.G. are co-first authors of this paper. C.Z. performed the FAST observations, data reduction, and stacking measurements.  H.G. conceived of the original idea and led the analysis. Y.G. processed all the DESI optical data. A.S. and X.Y. led the proposal for the DESI-HI programme. M.Z. and P.J. led the FASHI programme. Z.D. assisted with the implementation of the DESI fibre-assignment code in the mock catalogue. All co-authors contributed by their varied contributions to the science interpretation and data analysis. All co-authors contributed to the commenting on this manuscript as part of an internal review process. 

\section{Author Competing Interests}
The authors declare no competing interests.

\section{Data availability}
The data points for reproducing the figures are available from \url{https://zenodo.org/records/17790394}. The data we use in this article will be shared upon reasonable request to the corresponding author.

\section{Additional Information}
Correspondence and requests for materials should be addressed to Hong Guo (guohong@shao.ac.cn), Ming Zhu (mz@nao.cas.cn) and Peng Jiang (pjiang@bao.ac.cn). 

\clearpage

\end{document}